\documentclass[usedcolumn,useAMS,usenatbib]{mn2e}
\usepackage{times,rotating}
\usepackage{graphicx,subfigure,epsf}
\def\es{erg~s$^{-1}$}
\def\fluxu{erg~s$^{-1}$~cm$^{-2}$}

\def\xmm{{\it XMM-Newton\/}}
\def\cha{{\it Chandra\/}}
\def\etal{et al.\ }

\def\Lx {L_{\rm X}}

\def\rc {r_{\rm c}}

\def \h50 {$h_{50}$}
\def \h70 {$h_{70}$}

\def \nh {$N_{\rm H}$}
\def \rc {$r_{\rm c}$}
\def \mekal {{\sc mekal}}

\begin{document}

\title[High redshift FRIIs: large-scale X-ray environment]{High redshift FRII radio sources: large-scale X-ray environment}
\author[E. Belsole \etal]{E. Belsole$^{1,2}$\thanks{E-mail:
elena@ast.cam.ac.uk},  D. M. Worrall$^2$, M.J. Hardcastle$^3$, \& J. H. Croston$^3$\\
$^1$ Institute of Astronomy, University of Cambridge, Madingley Road, Cambridge, CB3 0HA, U.K.\\
$^2$ Department of Physics, University of Bristol, Tyndall Avenue,
Bristol BS8 1TL, U.K.\\
$^3$  School of Physics, Astronomy and Mathematics, University of Hertfordshire, College Lane, Hatfield, Hertfordshire AL10 9AB, U.K.}

\date{Accepted 2007 July 27. Received 2007 July 27; in original form 2007 May 19}
\maketitle

\begin{abstract}
We investigate the properties of the environment around 20 powerful
radio galaxies and quasars at redshifts between 0.45 and 1. Using \xmm\ and \cha\
observations we probe the spatial distribution and the temperature of
the cluster gas. We find that more than
60 per cent of powerful radio sources in the redshift range of our
sample lie in a cluster of X-ray luminosity greater than 10$^{44}$
\es, and all but one of the narrow-line radio galaxies, for which the
emission from the nucleus is obscured by a torus, lie in a cluster
environment. For broad-line quasars the X-ray
emission from the core dominates and it is more difficult to measure the cluster 
environment. However, within the statistical uncertainties we find no significant difference in
the properties of the environment as a function of the orientation to
the line of sight of the radio jet. This is in agreement with
unification schemes. Our results have important implications for
cluster surveys, as clusters around powerful radio sources tend to be excluded
from X-ray and Sunyaev-Zeldovich surveys of galaxy clusters, and thus can
introduce an important bias in the cluster luminosity function. Most of the radio sources are found close to pressure balance with the environment in which they lie, but the two low-excitation radio galaxies of the sample are observed to be under-pressured. This may be the first observational indication for the presence of non-radiative particles in the lobes of some powerful radio galaxies. 
We find that the clusters around radio sources in the redshift range of
our sample have a steeper entropy-temperature relation than local
clusters, and the slope is in agreement with the predictions of self-similar gravitational heating models for cluster gas infall. This suggests that 
selection by AGN finds systems less affected by AGN feedback than the local average. 
We speculate that this is because the AGN in our sample are
    sufficiently luminous and rare that their AGN activity is
    too recent to have caused the onset of measurable feedback and
    increased entropy in the clusters, especially in the cooler ones where locally the effect of feedback are expected to be most evident. If this is
confirmed by forthcoming X-ray missions it will improve our
understanding of the heating and cooling processes in high-redshift
galaxy clusters.
\end{abstract}


\section{Introduction}\label{intro}
\begin{table*}
\begin{center}
\caption{The sample}
\label{tab:sources}
\begin{tabular}{l|cclclcrr}
\hline
Source & RA(J2000) & Dec(J2000) & redshift  & scale &type & \nh & Literature\\
        & ($^{\rm h~m~s}$) & ($^{\circ~\prime~\prime\prime}$) & &
        (kpc/arcsec) & & ($\times10^{20}$ cm $^{-2}$) &  &\\
\hline
3C\,6.1 &  00 16 30.99 & +79 16 50.88  & 0.840 &7.63&  NLRG & 14.80 & this work\\  
3C\,184 &  07 39 24.31 & +70 23 10.74  & 0.994 &8.00&  NLRG & 3.45 & B04\\
3C\,200 &  08 27 25.44 & +29 18 46.51  & 0.458 &5.82&  LERG & 3.74 & this work\\
3C\,207 &  08 40 47.58 & +13 12 23.37  & 0.684 &7.08&  LDQ& 4.12 & Br02/G03\\
3C\,220.1& 09 32 39.65 & +79 06 31.53  & 0.610 &6.73&  NLRG & 1.87 & W01\\
3C\,228 &  09 50 10.70 & +14 20 00.07  & 0.552 &6.42&  NLRG & 3.18 & this work\\
3C\,254 &  11 14 38.71 & +40 37 20.29  & 0.734 &7.28&  LDQ& 1.90 & CF03/D03\\
3C\,263 &  11 39 57.03 & +65 47 49.47  & 0.646 &6.90&  LDQ& 1.18 & CF03/H02\\
3C\,265 &  11 45 28.99 & +31 33 49.43  & 0.811 &7.54&  NLRG & 1.90 & this work\\
3C\,275.1& 12 43 57.67 & +16 22 53.22  & 0.557 &6.40&  LDQ& 1.99 & CF03\\
3C\,280 &  12 56 57.85 & +47 20 20.30  & 0.996 &8.00&  NLRG & 1.13 & D03\\
3C\,292 &  13 50 41.95 & +64 29 35.40  & 0.713 &6.90&  NLRG & 2.17&  B04\\
3C\,295 & 14 11 20.65 & +52 12 09.04 & 0.461 & 5.530 &NLRG & 1.32 & A01\\
3C\,309.1& 14 59 07.60 & +71 40 19.89  & 0.904 &7.80&  GPS-Q& 2.30 & this work\\
3C\,330 &  16 09 34.71 & +65 56 37.40  & 0.549 &6.41&  NLRG & 2.81&  H02\\
3C\,334 &  16 20 21.85 & +17 36 23.12  & 0.555 &6.38&  LDQ & 4.24&  this work\\
3C\,345 &  16 42 58.80 & +39 48 36.85  & 0.594 &6.66&  CDQ&1.13 & G03\\
3C\,380 &  18 29 31.78 & +48 44 46.45  & 0.691 &7.11&  CDQ& 5.67& this work\\
3C\,427.1& 21 04 06.38 & +76 33 11.59  & 0.572 &6.49&  LERG&  10.90&  this work\\
3C\,454.3& 22 53 57.76 & +16 08 53.72  & 0.859 &7.68&  CDQ & 6.50 & this work\\
\hline
\end{tabular}
\vskip 10pt
\begin{minipage}{13cm}
Galactic column density is from \citet{dlnh}; NLRG means Narrow Line
Radio Galaxy; LERG means low-excitation radio galaxy, this is
  defined following \citet{JR97} as having [O~III]/H$_{\alpha}<0.2$
  and equivalent widths of [O III]$ < 3$ \AA;  LDQ means
  lobe-dominated quasar and CDQ means core-dominated quasar. The two
  classes of quasars are defined such that the ratio $R$ of 
  core to extended (lobe) flux density at 5 GHz (wherever available) on arc second scales is $R > 1$ for CDQ 
\citep[e.g.][]{gpcm93}. GPS indicates Gigahertz Peaked Spectrum sources. Redshifts and
positions are taken from \cite{3crrcat}.
 References in the last column are to papers searching for extended emission using data from the same observatory as used here: A01: Allen et al.
  2001; B04: Belsole et al. 2004; Br02: Brunetti et al. 2002; CF03:
  Crawford \& Fabian 2003; D03: Donahue et al. 2003; G03: Gambill et
  al. 2003 (this work does not find any cluster-like environment
  around the sources overlapping with our sample); H02: Hardcastle et al. 2002; W01: Worrall et al. 2001. 
\end{minipage}
\end{center}
\end{table*}

Much evidence supports the existence of a gaseous environment around
powerful radio galaxies and quasars. Theoretically, models of jet
confinement imply that an external medium with pressure similar to the
pressure found at the centre of galaxy clusters is required to keep
the jet collimated \citep{begelman84}. Observationally, two of the
most powerful radio galaxies, Cygnus A \citep{afejf84,rf96cyga} and
3C\,295 \citep[e.g.][]{hh86cyga,allen01-3c295}, are located in the centre of rich clusters
of galaxies. Studies based on optical observations find that, at
 redshift  $\sim 0.2-0.3$, powerful Fanaroff-Riley type II (FRII) sources
are found in relatively modest environments \citep[group scale, e.g.,][ and
references therein]{hl91,z97,woldetal00,best04}, while at higher
redshifts the environments of FRIIs may again be rich clusters
with richness comparable to 
Abell class I or higher \citep*[e.g.,][]{hg98}. This evidence
is provided mainly by studies based on galaxy
over-densities,
gravitational arcs \citep{deltorn97, wold02} and lensing shear of
surrounding field galaxies \citep{bs97}. Hence it has been apparent for
many years that powerful radio galaxies and
quasars should act as sign-posts of massive galaxy clusters at high redshift.

The clearest observational evidence for the existence of a cluster
around a radio source is the detection of an X-ray emitting, thermal,
large-scale medium surrounding it. Using {\it ROSAT} data, a number of
objects at $>0.5$ were inferred to lie in moderate to rich clusters
\citep[e.g.,][]{dmw94,crawford99, mjh99,mjh00}. However, the limited
sensitivity and resolution of the instruments on-board {\it ROSAT} did
not allow spectral confirmation of these results, and in some cases
emission from the central AGN was not separated from the extended
component.

X-ray studies of the cluster-like environment of radio sources are
complicated by the X-ray emission of other components, particularly
the AGN nucleus and radio lobes. The greatly improved spatial
resolution of the \cha\ observatory confirmed some of the clusters
detected with {\it ROSAT} \citep{dmw01,mjh02}
and found new clusters \citep[e.g.][]{siemigi05-3c186}. A few
observations with \xmm\ added to the cluster detections around radio
sources \citep{bel04}. However, studies based on small samples claim
that inverse Compton (IC) emission coincident with the radio lobes
dominates the X-ray extended emission associated with at least some
FRIIs at high redshift \citep*[e.g.,][]{ddh03, carilli02,mjh02},
obscuring most of the measurable thermal, cluster-like emission
\citep[e.g.,][]{acf03}. These studies thus need a precise modelling of
the Point-Spread-Function (PSF) and a precise measurement of the
non-thermal diffuse IC emission often found to coincide with the radio
lobes. What has been lacking until now is a uniform analysis of a
reasonably sized sample of objects aimed at characterising the cluster
X-ray emission, separating it from the emission from the nucleus and
radio lobes.

The cluster luminosity function can be used to place stringent
constraints on cosmological models 
\citep[e.g.,][]{evrard89,oukbla92,eke98,shuecker03}, and the detection of a
large number of clusters, in X-ray or via the Sunyaev-Zeldovich (SZ)
effect, is more efficient than measuring the redshift of millions of
galaxies, especially at high redshifts \citep[e.g.,][]{shueckeretal01}.
Although some high-redshift ($z>1$) galaxy clusters have been detected
lately as serendipitous sources \citep[e.g. ][ and references
therein]{mullis05, stanford06}, measurements of the typical gas mass and luminosity of clusters around active galaxies is lacking. This
has important implications because in the local universe 70 per
cent or more of massive clusters appear to harbour an active galaxy at
their centre, and at earlier times it is expected that galaxies are
more active, so that failing to include these objects in the
cluster luminosity function may result in an important bias
(\citealt{willott01, celfab04}; Belsole, Fabian, Erlund, in preparation). This bias may be a function of cosmic time since there are suggestions that the richness of radio-galaxy environments evolves with time, so that the detection and investigation of active galaxy environments 
will allow us to map the cosmic evolution of structure, which would be
hampered by a pure X-ray or optical selection.

The study of the X-ray cluster environments around radio sources is
  important in itself for furthering our understanding of accretion
  mechanisms and AGN feedback, particularly in comparison with local
  sources. At low redshift FRIIs and less powerful FRI radio galaxies
  seem to lie in different environments, with the latter preferring
  richer media \citep{mjh99}; the study of radio galaxies at
  different redshifts therefore allows us to shed light on the
  effect of the environment on the radio galaxy properties and the
  accretion mechanisms at various look-back times. Furthermore, a
  comparison between environments of radio sources with jets at
  different angles to the line of sight is a powerful way of testing
  orientation-based unification models.
 
In this paper we present the first systematic X-ray study of the
extended emission from powerful radio galaxies at $z>0.45$ to have been
carried out with the current generation of X-ray satellites, using a
moderately large sample of 20 radio galaxies and quasars from
the same catalogue. Of the 20 sources, 11 have previously
published studies of their extended X-ray emission; 5 of these were
studied by our group. Table \ref{tab:sources} gives references to the
literature for these objects. Clusters around 5 more sources are
detected in this paper for the first time. The X-ray properties of the nuclei of this sample
were discussed in \citet{bel06}. The lobe IC emission from most of the
sources was described in \citet{croston05b}. In this paper we report
the properties of the external environments around the sources in our
sample.

 Throughout the paper we use a cosmology with $H_{\rm 0}$ = 70 km
s$^{-1}$ Mpc$^{-1}$, $\Omega_{\rm m}$ = 0.3, $\Omega_{\Lambda}$ = 0.7.
If not otherwise stated, errors are quoted at the 1$\sigma$ confidence
level.

\section{ The sample}\label{sec:sample}
This study is based on the sample described in \citet*{bel06}. Sources
are taken from the 3CRR catalogue \citep*{3crrcat}, which is selected
on the basis of low-frequency (178-MHz) radio emission, and are in the
redshift range $0.45<z<1.0$.  In the present paper we have
added one source, 3C\,295, to the original sample. 3C\,295 was not
investigated by \citet*{bel06} as it was not part of the {\it Spitzer}
sample discussed in the above paper, although its nuclear X-ray
properties were discussed by \citet{hec06}. For consistency reasons we
include it in the present sample.
3C\,295 is a complex source in
the X-ray and the detailed analysis of its properties is not
comparable with the more global analysis that can be carried out
with the other sources in our sample, which is limited by photon
statistics. As a result, a detailed re-analysis of the source similar to the one carried out by \citet{allen01-3c295} is beyond the
scope of this paper. For the present work we limit ourselves to derive global properties from the 90 ks ACIS-I observation of 3C\,295, which, with an exposure 5 times longer, offers a much better statistics than the early ACIS-S observation used by Allen \etal This approach makes us confident that we are comparing this extreme source to the other sources in the sample by treating it in a similar way.

The sample is composed of a similar number of broad-line
quasars and narrow-line radio galaxies. Table~\ref{tab:sources} lists
the main properties of the sources including their quasar
classification. 

\begin{table*}
\caption{Observation log. Col 1: 3CRR name; Col. 2: Instrument used
for the observations, C stands for \cha, X for \xmm; Col.3:
Observation ID; Col. 4: Nominal exposure Time; Col. 5: Net exposure
time, after flare screening. For \xmm\ observations we give times for mos/pn detectors; Col. 6 pileup fraction.}
\label{tab:obslog}
\begin{tabular}{l|crclc }
\hline
Source & Instrument & Obs ID  & EXPOSURE & Screened time& pileup  \\
        &       &               & (ks) &  (ks) & \\
\hline
3C\,6.1 & C & 4363 & 20 & 20.0& 5\\  
      & C & 3009 & 36 & 35.7& 5\\  
3C\,184 & C & 3226 & 20 &18.9&1 \\
      & X & 0028540201&38.9 & 32/--& na\\
      & X & 0028540601& 40.9 & 26/16.4&na \\
3C\,200 & C & 838 & 16 &14.7& 1\\
3C\,207 & C & 2130 &  39& 37.5& 22\\
3C\,220.1&C & 839 & 21 & 18.5& 7\\
3C\,228 & C & 2453 & 12 & 10.6& 2\\
      & C & 2095 & 15.5& 13.8& 3\\
3C\,254 & C & 2209 & 31& 29.5& 20\\
3C\,263 & C & 2126 & 51& 48.8 & 26\\
3C\,265 & C & 2984 &59&50.6 & 1\\
3C\,275.1&C & 2096 &26& 24.8& 4\\
3C\,280 & C & 2210 & 63.5& 46.3& $<1$\\
3C\,292 & X & 0147540101 & 33.9 & 20/17&$<1$ \\
3C\,295 & X & 2254 & 92.1 & 90.9&0\\
3C\,309.1&C & 3105 & 17& 16.6& 6\\ 
3C\,330 & C & 2127 & 44& 43.8& 0\\
3C\,334 & C & 2097 & 33& 30.2& 9\\
3C\,345 & C & 2143 & 10& 9.0& 13\\
3C\,380 & C & 3124 & 5.5& 5.3& 16\\
3C\,427.1& C & 2194 & 39&39.0& 0\\
3C\,454.3& C& 3127 & 5.5& 5.5& 24\\
         & C& 4843 & 18.3& 18.0& 17\\
\hline
\end{tabular}
\end{table*}

Preparation of the \cha\ data is described in \citet*{bel06}. However for this paper we re-processed the data with {\sc ciao} 3.3.0.1 and {\sc caldb} v3.2.4, and added observation ID 4843 of 3C\,454.3 and ID 2254 of 3C\,295.  \xmm\ data preparation is described in \citet{bel04}, and we use the same data and results for this analysis. In Table \ref{tab:obslog} we give details of the X-ray observations used in this work.

\section{Spatial analysis}

\subsection{Imaging}

Diffuse, thermal X-ray emission associated with a cluster-like
environment is best detected at soft energies, between 0.5 and 2.5
keV, as this is the energy range where the \cha\ and \xmm\ mirrors are
the most sensitive. Thermal bremsstrahlung cluster radiation peaks in
this energy range, while the nuclear X-ray component and the lobe
inverse-Compton emission tends to have a harder spectrum. For this
reason we generated images of each source in the 0.5-2.0 keV (soft)
and 2.5-7.0 keV (hard) energy bands (we adopt a separation of 500
eV in order to avoid features associated with the Si edges, and
also to ensure that the two energy bands do not overlap). We then
applied a wavelet reconstruction algorithm provided by A.
Vikhlinin\footnote{http://hea-www.harvard.edu/RD/zhtools/} as an
efficient way to search for extended emission not associated with the
PSF, and at a detection limit of 4.5$\sigma$ above the background
level at the location of each pixel. In Appendix~\ref{app:A1} we show
for each source observed with {\it Chandra} 
the wavelet-decomposed and reconstructed
images in the soft and hard band. The same analysis for 3C\,184
and 3C\,292 was presented by \citet*{bel04} and so is not repeated
here. The radio emission at 1.4 GHz is shown by the superimposed
logarithmic contour. The comparison between the images at
soft and hard energy bands gives an immediate indication of the
presence of extended X-ray emission which may be associated with an
external environment of the radio source. For most of the sources,
extended X-ray emission seems to be associated with the radio lobes,
but many of the sources also show more symmetrical emission which
tends to fade at high energy.

\subsection{Radial profile}\label{sec:radprof}
\begin{table*}
\caption{Radial profile modelling results. Col.~1: source 3CRR
  name;  Col.~2: quality flag, where C implies a detection of
  extended emission with constrained structural parameters, D implies
  a detection with $F$-test probability for having an extended
  component $>98$ per cent but unconstrained structural parameters,
  and P implies a point-like source with no detected extended emission;
   Col.~3: best-fitting beta parameter;  Col.~4:
  best-fitting core radius;  Col. 5: number of counts predicted
  by the best fitting $\beta$-model included in a circle of radius
  corresponding to the detection radius R$_{\rm det.}$; Col.~6:
  external radius of the last annulus used to integrate the profile,
  this is also the radius of the inner circle of the annulus used for
  background subtraction;  Col.~7: number of counts predicted by
  the best fitting $\beta$-model out to a distance from the centre
  corresponding to $\rm R_{200}$; Col.~8: distance from the
  centre corresponding to the radius at which the density of the
  cluster is equal to an over-density of 200; Col.~9: Central
  surface brightness of the $\beta$-model;
  Col.~10: $\chi^2$/d.o.f. corresponding to the best-fitting
  $\beta$-model+PSF. If ``UL'' a 3$\sigma$ upper limit was calculated
  using appropriate values of $\beta$ and \rc; 
  Col.~11: $\chi^2$/d.o.f. corresponding to the best-fitting PSF model
  only; Col.~12: $F$-test null probability that the two-component
  model gives an improved fit for random data.}
\label{tab:betafit}
\begin{tabular}{l|clrrrrclrcr}
\hline
Source&	qual&	beta&$r_{\rm c}$&cnt$_{\rm det.}$& R$_{\rm det.}$&cnt($\rm R_{200}$)&$\rm R_{200}$& S$_0$&$\chi^2$/d.o.f.&$\chi^2$  & $F$-test prob.\\

      &     &       & (arcsec)  &                &(arcsec)       &                   & (kpc)      &(cts arcsec$^{-2})$ &  & (PSF fit) &(per cent)\\
\hline
3C\,6.1&	P&  	0.5f&	15.0f&	$<226$&		49.2&	$<235$&	620&	$<0.01$&	UL&	5.68/8&	 N/A\\
3C\,184$^{*}$&D&	0.66f&	20.0f&	$63\pm60$&  	87.5&   67$_{-25}^{+118}$&	790&	0.03$\pm0.02$& 0.6/5 &	3.13/6&	5.90e-3 \\
3C\,200&	C&	1.00$^{+0.20}_{-0.40}$&5.85$_{-3.49}^{+4.99}$&160$\pm47$&64.0&168$\pm50$&1140&2.31$\pm0.69$&2.20/3&13.57/6&0.105\\
3C\,207$^{**}$&	D&	0.80$_{-0.30}^{+0.20}$&6.1$_{-4.8}^{+3.9}$ &342$\pm43$&19.8&506$^{+958}_{-103}$&1146&3.46$\pm0.43$	&4.61/6	&67.60/9&6.76e-4\\
3C\,220.1&C&	0.58$^{+0.14}_{-0.08}$&3.24$_{-1.66}^{+2.94}$&852$\pm80$&39.4&1021$\pm96$&1150	&8.80$\pm0.82$&6.90/7&121.30/11& 1.89e-4\\
3C\,228&	D&     0.5f&15.0f&122$\pm38$&78.7&214$\pm140$&1072&0.05$\pm0.01$&5.81/7&16.37/8& 9.13e-3\\
3C\,254& 	C& 	0.65$_{-0.10}^{+0.15}$&1.95$_{-0.95}^{+1.50}$&252$\pm35$	&39.4&	273$^{+25}_{-14}$&568&10.7$\pm1.5$&10.64/7&62.50/10&4.41e-3\\
3C\,263& 	P&	0.67f&	10f&	$<904$&	64.0	&$<968$&1173	&$<1.66$&UL&86.50/12& N/A\\
3C\,265&	D&	1.00$_{-0.33}^{+1.00}$&2.85$^{+3.15}_{-1.85}$&53$\pm17$&24.6&56$\pm18$&436&3.3$\pm1.00$&2.54/2&12.03/5&0.299\\
3C\,275.1&P&	0.60f&	7.0f&	$<113$&	19.7	&$<129$	&494	&$<0.34$&UL&7.54/7&N/A\\	
3C\,280&	D&	0.50f&	8.0f&	$98\pm31$&32.0&$213^{+300}_{-153}$&119& 0.15$\pm0.05$&	2.58/4& 12.86/5& 1.62e-2\\
3C\,292$^{*}$&C&	0.80$_{-0.25}^{+0.50}$&19.7$^{+26.8}_{-14.0}$&523$\pm66$&100.0&528$^{+111}_{-68}$& 707&0.40$\pm0.05$&1.62/7	&65.08/10& 5.61e-6\\
3C\,295 & C & 0.52$^{+0.01}_{-0.01}$ & 3.4$^{+0.25}_{-0.25}$&
11400$\pm 75$ & 108.2 & 13130$\pm 160$ & 1268 & $55.8_{-3.4}^{+2.9}$&
84/55 &3.9$\times 10^5/58$ & 0 \\
3C\,309.1&C&	0.56$_{-0.04}^{+0.19}$&1.78$_{-1.33}^{+4.24}$&141$\pm26$&39.4&	164$_{-42}^{+122}$&366&3.76$\pm0.68$&0.87/3&29.40/6&8.57e-3\\
3C\,330&	C&	0.65$_{-0.15}^{+0.55}$&7.40$_{-5.30}^{+12.1}$&158$\pm26$&39.4&166$^{+46}_{-37}$&644&0.48$\pm0.08$&1.14/3&37.35/6&8.97e-3\\
3C\,334&	P&	0.67f&	10.0f&	$<155$&	35.4	&$<191$&496	&$<0.35$&UL&3.54/8&N/A\\
3C\,345&	P&	0.67f&	15.0f&	$<127$&	39.4	&$<154$&717	&$<0.13$&UL&6.93/11&N/A\\	
3C\,380&	P&	0.50f&	7.0f&	$<105$&	88.6	&$<109$&677	&$<0.13$&UL&15.98/9&N/A\\
3C\,427.1	&C&	0.50$_{-0.10}^{+0.05}$&5.85$_{-2.85}^{+3.60}$&305$\pm37$&49.2&	894$^{+26}_{-444}$&1315	&0.59$\pm0.07$&1.81/4&69.58/7&1.26e-3\\
3C\,454.3&P&	0.50f&	5.0f&	$<385$&	29.5	&$<480$&413	&$<1.27$&UL&14.47/9&N/A\\
\hline
\end{tabular}
\vskip 8pt
\begin{minipage}{500pt}
(*) from XMM analysis. The listed values refer to the PN camera only. See \citet{bel04} for details.

(**) although the $\beta$-model parameters are constrained we were not able to constrain the temperature of the extended emission. We thus decided to be conservative and give a quality flag D instead of C for this source (see discussion in the text). 

\end{minipage}
\end{table*}

We carried out a more quantitative characterisation of the spatially
extended emission by extracting a radial profile of each source. The
profile was centred on the emission peak of each source, and point
sources other than the central AGN were excluded. Here we focus on
extended emission associated with the external environment and not the
extended non-thermal X-ray emission associated with the lobes
\citep{croston05b}. Although in Croston \etal\ we showed that many of
the sources in the sample contribute very few counts to the lobe
emission, to be conservative we excluded the spatial regions
coincident with the radio lobes before extracting the radial profile.
The masked area was properly taken into account when deriving physical
parameters. Any emission at radio frequencies from the CDQs --
not discussed in \citet{croston05b} -- is within 5 arcsec of the
centre of the source and so not relevant to the analysis
discussed here. For details of the detection of X-ray
counterparts of three of the sources see \citet{sambruna02}
(3C\,345), \citet{marshall05} (3C\,380), and
\citet{marshall05,tavecchio07} (3C\,454.3).

 We used the energy band 0.5-2.5 keV which optimises the instrument sensitivity
to the soft thermal emission over the AGN and particle contributions.
Exposure corrections were applied using a map calculated at the peak
energy of the global spectrum, which was found to be at 1 keV for most
of the sources. When multiple exposures were available, the radial
profile was extracted from a mosaic event list and the mosaic exposure
map was used. The background region was chosen by selecting an annular
region between the distance from the centre at which the radial
profile becomes flat and at which it increases because
particle-dominated background has been erroneously corrected for
vignetting. The inner radius of the background annulus is
specific to each source (see Table \ref{tab:betafit}). We limited our
analysis to the S3 chip in all cases but 3C\,295 (ACIS-I), and the background area was also
selected from this chip.

To assert the presence of extended emission, modelling of the PSF is
crucial, especially in the context of these sources which harbour
bright AGN. To obtain the best representation of the central point
source emission we generated a model of the PSF at the off-axis
position of each source using the Chandra Ray Tracer (ChaRT) and
MARX\footnote{http://cxc.harvard.edu/chart/}. We used the option of
giving the spectrum of the source as the input parameter in order to
generate a PSF model appropriate for the spectral distribution. The
output file of ChaRT was then used to generate an event list of the
PSF using MARX. We then extracted a radial profile of the PSF for each
source and used this model to fit the source radial profile. We adopted an iterative process as the PSF model is
dependent on the blurring parameter which is an input to MARX. We changed
this parameter in order for the PSF to fit the first two bins of the
radial profile of the source. For sources with a pileup fraction of
more than 10 per cent we used MARX to generate a piled-up PSF model and
we used this event list as the best representation of the point
source. Radial profiles were initially fitted with a PSF. When this
was not an acceptable fit, we added a $\beta$-model. We varied the 
$\beta$ parameter and core-radius, $r_{\rm c}$, when the photon statistics were
sufficient to carry out a two-parameter fit. In all other cases we
fixed the $\beta$ and \rc\ parameters to values (generally 2/3,
and 120-150 kpc) commonly observed in local clusters, allowing for the
possibility that clusters at high redshift are more concentrated
(i.e. smaller core radii and  $\beta$ than canonical values in the
local universe should not be seen as unusual). The
results of the radial profile fitting are listed in Table
\ref{tab:betafit}, and the individual fits can be inspected in
Appendix~\ref{app:A1} for the sources observed with \cha, while for 3C\,184 and 3C\,292, observed with \xmm, the profiles are shown in \citet{bel04}.

\begin{figure}
\begin{centering}
{\epsfxsize=3.3in \epsffile{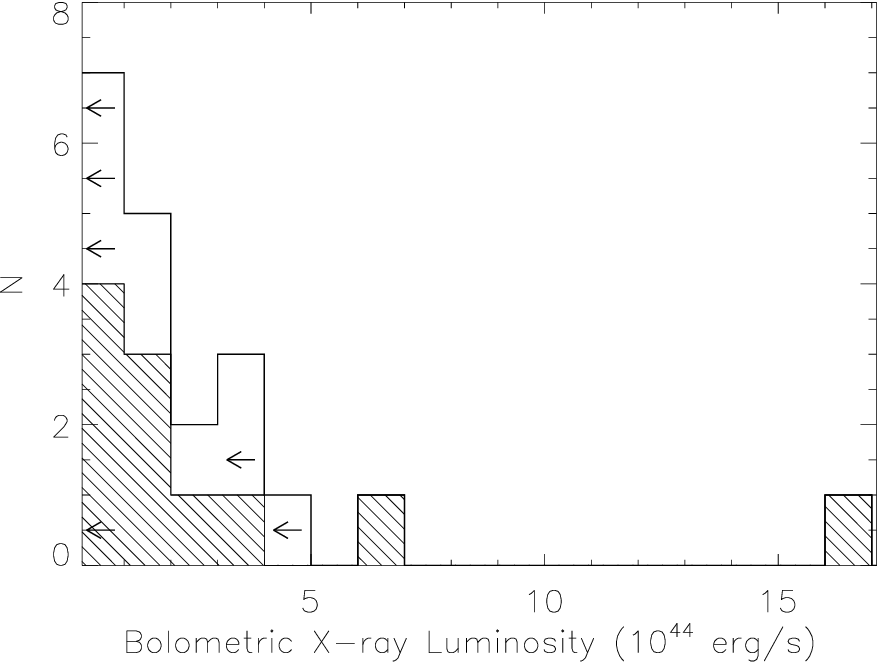}}
\caption{Bolometric X-ray luminosity distribution. The shaded area corresponds to the radio galaxies sub-sample}
\label{fig:LXhisto}
\end{centering}
\end{figure}

\begin{figure*}
\begin{centering}
\includegraphics[scale=0.6,angle=0,keepaspectratio]{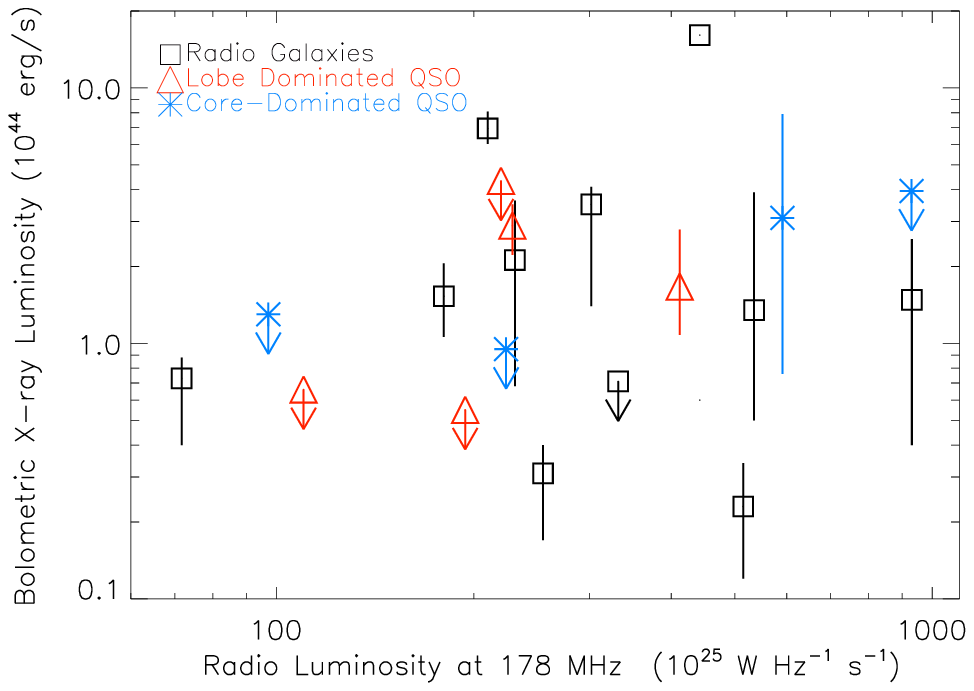}
\includegraphics[scale=0.6,angle=0,keepaspectratio]{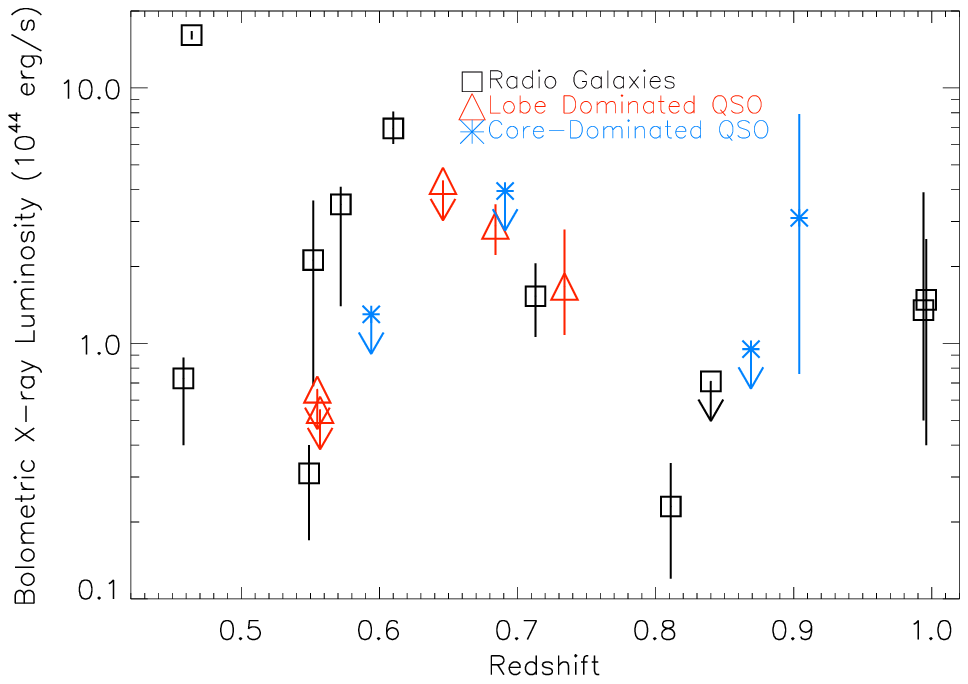}
\includegraphics[scale=0.6,angle=0,keepaspectratio]{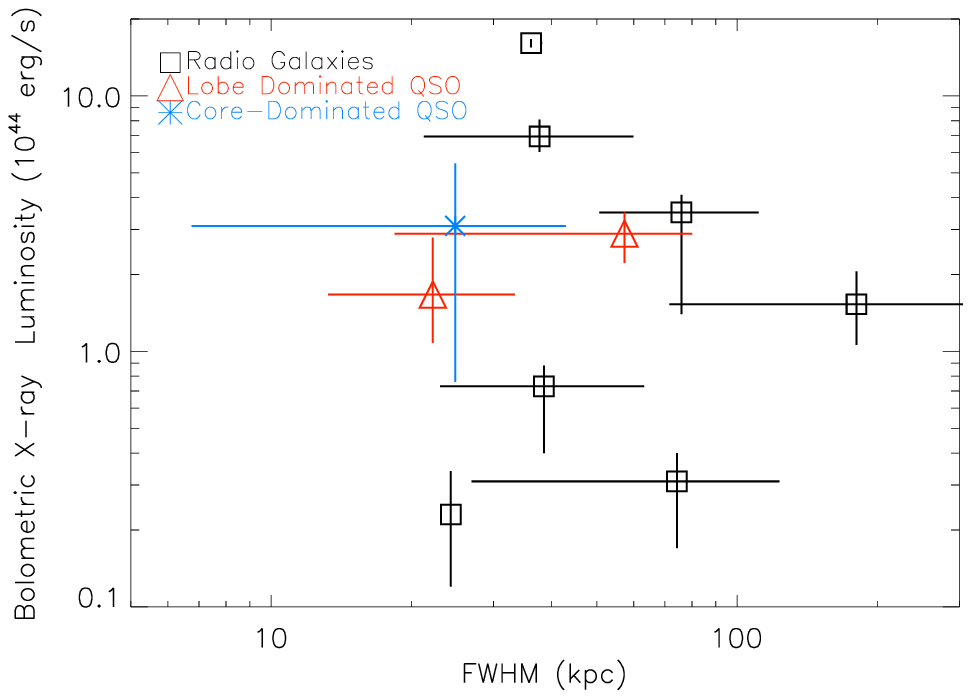}
\caption{Top: Bolometric X-ray luminosity versus radio power of each galaxy as measured at 178 MHz. Middle: Bolometric X-ray luminosity versus redshift.  Bottom: Bolometric X-ray luminosity versus Full-Width-Half-Maximum ($r_{FWHM}$) defined as in Eq. 2. Only sources for which fitting of $\beta$ and $r_{\rm c}$ were possible are used for this plot.}
\label{fig:rel1}
\end{centering}
\end{figure*}

\section{Spectral analysis}
\begin{table*}
\caption{Spectral analysis: Col.~1: source 3CRR name; 
Col.~2: radius of the circle used to integrate the spectrum. This may
be smaller than the outer circle used to extract the radial profile;
 Col.~3: number of net counts in the energy range 0.5-2.5 keV
contained in the area used for the spectral analysis. This is lower
than the number of counts in the radial profile as the area masked
before extracting the profile is larger to account for the point
source emission. When two values are listed it means that two
exposures have been used. For 3C\,184 and 3C\,292 we list the counts
from the whole EPIC;  Col.~4: best-fit mekal model temperature
and 1 $\sigma$ errors for one interesting parameter; Col.~5:
chemical abundances used for the fit. The \citet{grsa} table for solar
abundances was adopted and all values were fixed; Col.~6:
Normalisation of the \mekal\ model;  Col.~7: $\chi^2$/d.o.f.
corresponding to the best-fitting model; Col. 8: unabsorbed
X-ray flux in the energy range 0.5-2.5 keV; Col.~9: unabsorbed
X-ray luminosity of the thermal component in the energy range 0.5-2.5
keV; Col.~10: total (bolometric) X-ray luminosity of the thermal
component within the detection radius of the spectrum; Col.~11:
Bolometric X-ray luminosity. This was calculated by extrapolating the
number of counts out to the virial radius using the radial profile
distribution and using the count rate to adjust the normalisation of
the spectral model to obtain the bolometric X-ray luminosity. Errors
are the quadratic sum of the statistical errors derived from the
spectral and radial profile analyses.}
\label{tab:spfit}
\begin{tabular}{l|rclcllllll}
\hline
Source&	R	&cnts  & k$T$& $Z/Z_{\odot}$&	Norm & 	$\chi^2$/dof &  $f_{\rm X}$ ($\times10^{-14}$& L$_{\rm X}(\times10^{43}$&L$_{\rm X}^{Bol_{sp}}$ &L$_{\rm X}^{Bol}(\times10^{44}$ 	\\	
        & (arcsec)   &                & (keV) &               &    ($\times10^{9}$ cm$^{-5}$)         & &  \fluxu) & \es) &  ($\times10^{44}$ &  \es) \\
&          & 0.5-2.5 keV    &     &               &              & & 0.5-2.5 keV   & 0.5-2.5 keV& \es)         & Extrapol.                    \\

\hline
3C6.1&	49.2&	60+14&		2.0f&0.5&$<3.24$&UL&$<0.72$&$<2.1$&$<0.59$&$<0.71$\\
3C184$^{a}$&40.0&201$^b$&         3.6$_{-1.9}^{+10.7}$&0.3&2.27$_{-0.46}^{+0.53}$&43.13/47&0.5$_{-0.10}^{+0.12}$&2.1&0.5&1.35$_{-0.90}^{+2.55}$\\
3C200&	24.6&	79&		3.91$_{-1.82}^{+7.67}$&0.3&3.40$_{-0.69}^{+0.63}$&7.27/10$^C$&1.26$_{-0.52}^{+0.24}$&0.1& 0.26& 0.73$_{-0.33}^{+0.15}$\\
3C207&	19.8&	127$^{*}$&	5.0f&            0.5&2.29$_{-0.47}^{+0.47}$&5.33/5& 0.8$_{-0.2}^{+0.2}$&1.4&0.38&2.89$_{-0.67}^{+0.61}$\\
3C220.1&24.6& 	338 &		4.65$_{-0.89}^{+1.29}$& 0.3&16.38$_{-1.12}^{+1.18}$&10.86/13&5.2$_{-0.34}^{+0.39}$&6.9&1.86& 6.94$_{-0.90}^{+1.14}$\\
3C228&	49.2&	66+66&		3.87$_{-1.51}^{+6.86}$&0.3&7.05$_{-1.02}^{+1.09}$&4.86/12&2.3$_{-0.40}^{+0.36}$&2.5&0.62& 2.12$_{-1.44}^{+1.50}$\\
3C254&	34.4&	82$^{*}$&		1.54$_{-0.87}^{+7.06}$&	0.5&1.33$_{-0.43}^{+0.87}$&1.62/3& 0.34$_{-0.11}^{+0.22}$&0.9&0.18&1.67$_{-0.59}^{+1.12}$\\
3C263&	64.0&	162$^b$&	5.0f	&       0.3&$<2.9$& UL& $<0.9$	&1.4& 0.38& $<4.33$\\
3C265&	24.6&	16&		1.05$_{-0.46}^{+0.34}$&0.5& 0.55$_{-0.25}^{+0.23}$& 4.73/5$^C$& 0.14$_{-0.06}^{+0.06}$& 0.6& 0.10&0.23$_{-0.11}^{+0.11}$\\
3C275.1&19.7&	8&		1.0f	&       0.3&$<0.86$&UL&$<0.35$&0.5&0.09&$<0.55$\\
3C280&	20.0&	53&		5.0f	&	0.3&1.52$_{-0.39}^{+0.34}$&0.41/3$^C$&0.34$_{-0.09}^{+0.07}$&1.3&0.40&1.48$_{-1.08}^{+1.08}$\\
3C292$^{a}$&100&310&            2.18$_{-0.83}^{+3.12}$.&0.3&5.85$_{-1.62}^{+1.57}$&10.58/12&1.47$_{-0.41}^{+0.41}$&3.11&0.64&1.53$_{-0.47}^{+0.53}$\\
3C\,295 & 108.2 & 8034 & $4.74_{-0.23}^{+0.26}$  &
0.48$\pm0.08$& 133$\pm5$&163/190& 52$\pm2$& 36&
9.86&$16.1 \pm 0.6$ \\ 
3C309.1&32.0&	31$^{*}$&	0.87$_{-0.07}^{+0.86}$&0.7&1.11$_{-0.79}^{+1.52}$&0.34/4$^C$&0.24$_{-0.17}^{+0.33}$&1.74&0.29&3.1$_{-2.34}^{+4.80}$\\
3C330&	39.4&	146&		1.59$_{-0.63}^{+3.78}$&0.2&2.57$_{-0.97}^{+0.30}$&1.48/6$^C$&0.63$_{-0.24}^{+0.07}$&	0.82&0.17&0.31$_{-0.14}^{+0.09}$\\
3C334&	35.4&	24&		1.0f&	        0.5&$<2.51$& UL&$<1.04$&0.15&0.25&$<0.66$\\
3C345&	39.4&	37$^{*}$&	2.0f&		0.5&$<4.43$& UL&$<1.15$&0.20&0.40&$<1.3$\\
3C380&	88.6&	76$^{*}$&	2.0f&		0.5&$<11.7$& UL&$<3.11$&6.8&1.36&$<3.95$\\
3C427.1&24.6&	168&		5.66$_{-2.35}^{+9.59}$&0.3&3.39$_{-0.57}^{+0.60}$&3.94/3&1.2$_{-0.20}^{+0.20}$&1.27&0.38&3.5$_{-2.10}^{+0.60}$\\
3C454.3&29.5&	77+31$^{*}$&	1.0f&		0.5&$<4.2$&UL&$<1.0$&5.5&0.96&$<0.95$\\
\hline
\end{tabular}
\vskip 8pt
\begin{minipage}{500pt}
$^*$ counts are partially from the wings of the PSF and have been modelled before extracting the thermal model values.

$^a$ See \citet{bel04} for details. For 3C\,184 the number of counts and the $\chi^2$ correspond to the whole spectral region (nucleus not excluded) and fitted model. 
For 3C\,292, the area used for the spectral analysis was an ellipse on major and minor semi-axis equal to 101 and 64 arcsec respectively. A sector in the region corresponding to the radio lobes was also excluded. 

$^b$ 3C263: All counts are from a point source; 3C\,184: part of the counts are from the point source as it cannot be resolved with XMM data  

$^C$ C-statistics were used, and the values quoted are the C-statistic parameter and the Pulse Height Amplitude (PHA).

\end{minipage}
\end{table*}

We selected the spatial region for spectral extraction on the
basis of the shape of the radial profile. We initially extracted a background-subtracted 
spectrum of the source in the same area as used for the radial profile
analysis, but we excluded the nuclear region with circles of radii 2-4
arcsec for the non-piled up sources (weak core or pileup fraction $< 10$) and up to 10 arcsec
when pileup was significant (strong cores, pileup fraction $> 10$). We also masked the areas
coincident with the radio lobes as detailed in the previous section.
For those sources with no X-ray lobe detections or upper limits on the
X-ray emission from the lobes \citep{croston05b}, we did not mask the
radio lobe area. In the case of CDQs the radio emission is within the area used to mask the core.

We initially fitted all spectra with a {\sc mekal} model with fixed
chemical abundance \citep[using the solar abundances of][]{grsa}. For
a number of sources it was necessary to add a power-law model to
account for residuals of the PSF. In these cases, to account for the
flattening of the power-law spectrum in the wings of the PSF the
power-law index was fixed to be at the lower limit of the best-fit
spectral index obtained for the core spectrum of the source \citep*{bel06} and
we adjusted the normalization of the power-law model to be appropriate for the PSF
fraction in the area used for the cluster spectrum. For those sources
with no-detected extended thermal emission we calculated $3\sigma$
upper limits for a thermal component in addition to the power law. The
temperature in this case was fixed between 1 and 5 keV depending on
the shape of the power law spectrum.

Results of the spectral analysis are summarised in Table~\ref{tab:spfit}.

\section{How luminous are radio galaxy clusters ?}
The bolometric X-ray luminosity was calculated for each source adopting the spectral best-fit temperature (the fixed value was used when upper limits are considered) and extrapolating the spectrum to the whole energy range, and the $\beta$-model out to the virial radius, here assumed to be $R_{\rm 200}$.  This was calculated as:
\begin{equation}
R_{200} = h(z)^{-1}\times B_{200}\times(kT/5~ {\rm keV})^{\beta}
\end{equation}

\noindent In the expression above,  $B_{200}$ and $\beta$ are taken
      from the $R-T$ relation of \citet{arnaud05}, using
     the best-fit or adopted temperature of the cluster from Table 4, and 
 $h^2(z) = \Omega_m (1+z)^3 + \Omega_{\Lambda}$.

Uncertainties on the bolometric $L_{\rm X}$ were computed by adding in
quadrature the statistical fractional error on the normalisation of the
spectrum and the relative fractional error on the number of counts
predicted by the radial-profile best-fit model.
Figure~\ref{fig:LXhisto} shows the distribution of the bolometric
X-ray luminosity for the whole sample; the radio galaxy subsample
is shown by the shaded area. Upper limits are also included in the
histogram.

More than 60 per cent of the radio sources in the sample are found to
lie in environments of X-ray luminosity greater than 10$^{44}$ erg
s$^{-1}$, and more than 30 per cent lie in clusters of luminosity
greater than $2\times10^{44}$ erg s$^{-1}$. For 6 out of 9 quasars we were
unable to measure the $L_{\rm X}$ and only an upper limit indicates
the brightness of the cluster. On the other hand, only one RG,
3C\,6.1, does not show detectable extended emission. We believe that
this effect is mostly due to observational constraints: the quasars of
the sample have bright cores and they have been targeted mainly to
study the central bright source. As a result, the observation exposure
times are sometimes lower than those of the RGs. The bright core
is difficult to model, especially if it is piled up, so that
extended emission may be simply hidden, rather than nonexistent.
We cannot rule out the possibility that quasars at high redshift may inhabit less rich environments than radio galaxies
 as 
most of the quasars upper limits are in the low-luminosity part of the $L_{\rm X}$ distribution.  

In the top panel of Figure~\ref{fig:rel1} we show the distribution of the
bolometric X-ray luminosity as a function of the isotropic radio
luminosity as measured at 178 MHz. We find no significant correlation
between the two quantities for either the radio galaxies and the quasars.

We also find no significant correlation between the bolometric X-ray luminosity
and  redshift (Figure~\ref{fig:rel1}, middle), although a peak may
be present around $z=0.6$. To explore the evolution of the
luminosity function of clusters around powerful radio-loud active
galaxies it would be necessary to compare the objects drawn from this
sample with objects from a broader redshift range; this would allow us
to search for correlations between X-ray luminosity and redshift.

We also looked for a correlation between the extent of the gas
distribution and the X-ray luminosity. The full-width-half-maximum
(FWHM) of the $\beta$ model describing the intra-cluster medium (ICM) distribution is

\begin{equation}
r_{\rm FWHM} = 2~r_{\rm c} \times (0.5^{\frac{2}{1-6\beta}}-1)^{1/2}
\end{equation}

\noindent where $r_{\rm c}$ is the core radius. 
Figure ~\ref{fig:rel1}, bottom, shows that for the objects for
which fitting of the radial profile was possible there is 
no statistically significant correlation between the FWHM of the gas
        distribution and its bolometric luminosity. However, the radio
galaxies are preferentially found towards the right side of the plot, and there is a trend of an inverse correlation between the two quantities.
This result may be an indication that radio galaxies tend to lie in
more spatially extended clusters, although this does not imply
richer environments, as RGs occupy the whole $L_{\rm X}$ range in the
plot. Speculatively, this plot may suggest either that
more luminous clusters are more concentrated, or that, for the most
luminous clusters, we are only detecting their central, denser and
thus more luminous part.

\section{The state of the radio source}
\begin{table*}
\caption{Physical parameters. Col.~1: source 3CRR name;  Col.~2: Central proton density;  Col.~3: Central pressure of the external medium; Col.~4: average distance of the radio lobe from the centre of the source (and cluster). This is the distance used to calculate the external pressure at the location of the lobe;  Col. 5: External pressure of the environment at the location of the lobe. This should be considered as an average value;  Col.~6: IC minimum (equipartition) internal pressure  Col.~7: IC internal pressure when it was possible to measure it from X-ray observations (see Croston et al. 2005b)}
\label{tab:phcond}
\begin{tabular}{l|clrlccc}
\hline
Source    & n$_0$	    &P$_0$      &Rlobe & P$_{ext}$ &  P$^{min}_{int}$ & P$_{IC}$ \\
          & ($\times10^{-2}$ cm$^{-3}$)             & ($\times10^{-12}$ Pa)                    &(arcsec)& ($\times10^{-12}$Pa)                    &  ($\times10^{-12}$ Pa)                          &($\times10^{-12}$ Pa)\\
\hline
\hline
3C\,6.1   &$<0.6$	            &$<4.3$                   &8.4   &	$<3.5$                 & 	3.1 &	-- \\
3C\,184   &0.5$_{-0.2}^{+0.3}$	    &6.4$_{-2.5}^{+4.2}$      &2.5   &	$6.3^{+4.1}_{-2.5}$    &       72.0 &384.0$\pm115.0$ \\
3C\,200   &4.8$_{-2.6}^{+3.5}$      &68.45$_{-36.53}^{+49.35}$&9.5   &	$9.9^{+1.8}_{-3.6}$    &	0.5 &0.58$\pm0.07$ \\
3C\,207   &3.8$_{-1.3}^{+13.0}$     &68.50$_{-24.31}^{+238·0}$&4.8   &	$38.5^{+3.5}_{-6.8}$   &	2.7 &3.9$\pm0.6$ \\
3C\,220.1   &10.8$_{-5.8}^{+20.0}$    &183.0$_{-85.0}^{+157.0}$ &12.0  &	$17.6^{+1.8}_{-1.6}$   &	0.8 &   -- \\
3C\,228   &1.1$_{-0.9}^{+10.3}$     &15.15$_{-12.9}^{+145.9}$ &18    &	$3.5^{+0.8}_{-1.5}$    &	1.0 &	--\\
3C\,254   &16.9$_{-7.8}^{+13.9}$    &95.3$_{-44.0}^{+78.2}$   &7.5   &	$6.5^{+0.5}_{-0.5}$    &	4.0 &	--\\
3C\,263   &$<2.4$	            &$<44.0$                  &	18.0 &	$<10.0$                &	1.0 &2.6$\pm0.5$  \\
3C\,265   &6.7$_{-4.0}^{+13.9}$     &25.7$_{-15.2}^{+53.2}$   &	19.5 &$0.08^{+0.2}_{-0.05}$    &	0.8 &0.9$\pm0.1$  \\
3C\,275.1 &$<1.1$	            &$<4.04$                  &	5.8  &$<2.5$                   &	2.1 &2.5$\pm0.3$  \\
3C\,280   &$<0.9$	            &$<16.18$                 & 5.5  &$<12.1$                  &	8.0 &12.3$\pm3.1$  \\
3C\,292   &0.9$_{-0.4}^{+0.7}$      &6.8$_{-2.9}^{+5.9}$      &	37.5 &$1.1^{+0.2}_{-0.1}$     &	0.8 &   --  \\
3C\,295 & $12.7 \pm 0.6$ &$220 \pm 10$&2.0&$170 \pm 10$&100&--\\
3C\,309.1 &13.53$_{-10.2}^{+135.2}$ &43.5$_{-33.2}^{+435.3}$  &	--   &--     &	-- &   -- & \\
3C\,330   &1.3$_{-0.7}^{+3.0}$      &7.7$_{-3.9}^{+17.7}$     &	18.0 &	$1.2^{+0.2}_{-0.2}$    &	1.7 &2.3$\pm0.4$  \\
3C\,334   &$<1.0$	            &$<3.63$                  &	16.1 &	$<4.9$                 &	0.7 &0.9$\pm0.4$  \\	
3C\,345   &$0.9$                    &$<6.5$                   & --  & --  & -- & --  \\
3C\,380   &$1.3$                    &$<9.2$                   & --  & --  & -- & --   \\
3C\,427.1   &2.2$_{-1.3}^{+0.6}$      &45.0$_{-26.0}^{+13.0}$   &	7.5  &	$13.8^{+2.3}_{-2.1}$&	3.3 &3.7$\pm0.6$  \\
3C\,454.3 &$<4.9$                   &$<18.1$                  & --  & --  & -- & --   \\
\hline
\end{tabular}
\vskip 8pt
\end{table*}

\begin{figure}
\begin{center}
\includegraphics[scale=0.5,angle=0,keepaspectratio]{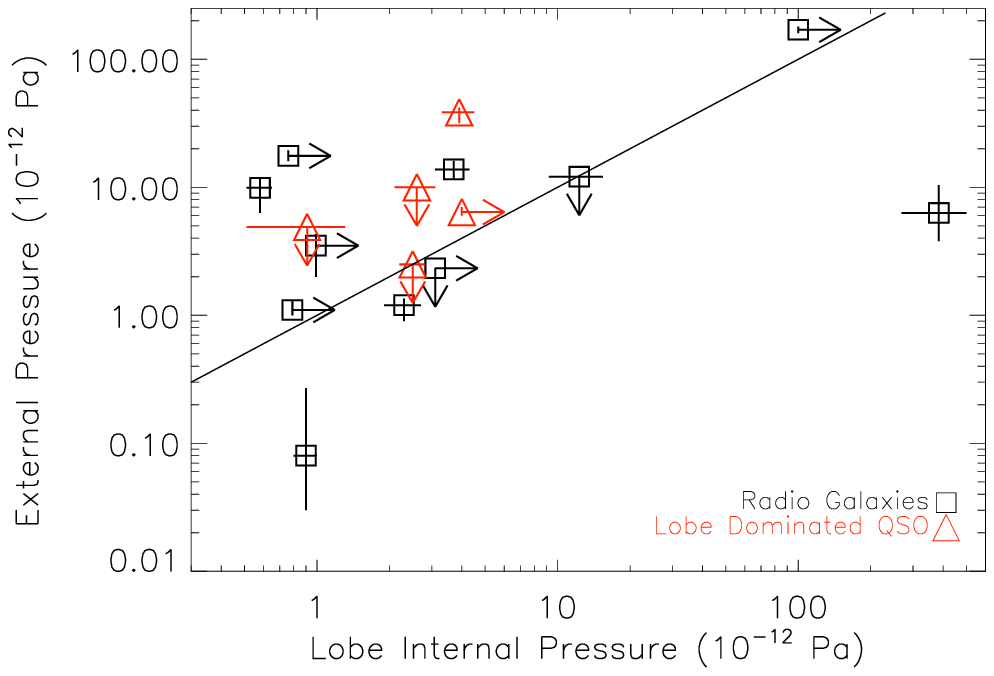}
\caption{External (cluster) pressure at a position corresponding to
the average distance of the radio lobe from the cluster centre versus total internal pressure calculated from an X-ray component of CMB inverse-Compton scattered emission in the lobe where measured, or the minimum pressure (Croston et al. 2005b). The continuous line indicates equality between external and internal pressure.}
\label{fig:pvsp}
\end{center}
\end{figure}

We computed the central density, central pressure and the pressure at
the location of the lobe using the best fit $\beta$ model and the
temperature that characterise the sources in the sample. Whenever
only upper limits of the gas distribution were available, we
calculated 3$\sigma$ upper limits for central density and pressure. We
used the relation given in \citet{bw93} to derive central proton
density and pressure from the central count density. Results are
tabulated in Table~\ref{tab:phcond}. The pressure of the external
environment was calculated at (or at times extrapolated to) a distance
corresponding to half the distance between the centre and the edge of
the radio lobe (sometimes terminating in a hot spot). This should
account for the fact that radio lobes have a cylindrical shape so the 
pressure at
the cylinder edges is a function of the gas pressure and so the
distance from the centre of the cluster.

The internal pressure of the radio source was calculated by combining
X-ray (IC) and radio data for those sources with detected X-ray
emission from the radio lobes \citep[][ for details]{croston05b}. For all other sources minimum pressures
were computed using the radio data only. The IC and minimum
internal pressures were obtained using the code of \citet{mjh98b},
using the radio and X-ray flux measurements, lobe volumes, and
electron energy model parameters for each source given in
\citet{croston05b}. For CDQs we are unable to
calculate the radio source pressure as the jet dominates the X-ray and
radio emission (see Sec.~\ref{sec:radprof}) and they are thus ignored in this analysis. In
Figure~\ref{fig:pvsp} we plot the external, thermal gas pressure versus
the internal pressure of the radio lobes. Minimum pressures are
treated here as lower limits. The continuous line shows the equality
of internal and thermal pressure.

Most of the sources appear to be close to pressure balance, with
only a few exceptions. The source at the far right edge of the plot is
3C\,184 whose size is very small (only few arcsec). We have
argued in \cite{bel04} that the size and relative pressure of
this source suggest that it may be in a young phase and beginning
to expand into the external medium ( see also \citet{siemigi05-3c186} for a similar conclusion about 3C\,186). 3C\,265 also has lobes that are
over-pressured with respect to the external gas. The sources
that appear to have lobe pressures that are significantly lower
than the pressure of the ICM are 3C\,200, 3C\,207, 3C\,220.1 and
3C\,427.1. 
The radio lobe pressure of 3C\,220.1 is a lower limit (minimum pressure) so
the real pressure can be higher and in agreement with pressure equilibrium. 
3C\,207 has a bright core and caution should apply as the
measurements of external pressure may have a larger than statistical
error due to additional contamination of the extended environment by
the point source. 
The remaining sources are the two low-excitation RGs
(LERG) in the sample and their
lobes appear genuinely under-pressured with respect to their environments. 
As lobes cannot
evolve to a state where they are truly underpressured, this suggests
that our pressure estimates for these sources may be incorrect.

Earlier evidence has suggested that LERGs prefer richer
environments than high-excitation objects
\citep[e.g.,][]{mjh04,reynolds05}. 
Our result suggest that even at high redshift, powerful LERGs
appear to inhabit rich environments. 
The behaviour of their
nuclei, both from an optical/infrared and X-ray perspective, appears
more similar to that of less powerful FRI radio galaxies, which
are almost all LERGs. The X-ray properties of the cores of  3C\,200 and 3C\,427.1 \citep*{bel06} are definitively consistent with this picture.

It is well known that the minimum pressures in the lobes of
low-power FRI sources are almost always found to be lower, often by an
order of magnitude or more, than the external pressures estimated by
observations of hot gas \citep[e.g.][]{croston03,dunn06}, requiring
either significant departures from equipartition in the lobes or, more
likely, a dominant contribution to the lobe pressure from
non-radiating particles. Our results point to the intriguing
possibility that some high-power LERGs may resemble low-power sources
in this respect as they do in their nuclear properties. This would
imply that in fact the work done by the radio source on the
environment is higher than that measured from radio observations alone
(i.e. minimum energy). Thus, in contrast with what we and
others have found to be the case for narrow-line radio galaxies and
quasars, for the two LERGs in our sample we have
measurements of the IC from X-ray observations \citep{croston05b}. We
still observe the sources to be over-pressured by their environments,
even though the measure we have of their energy density is more
realistic than minimum energy.

There is evidence that some other LERG FRII sources have
minimum lobe pressures that lie substantially below the external
thermal pressures (e.g. 3C\,388, \citealt{kra06}; 3C\,438,
\citealt{kraft07b}) but this is the first time that we have been able to
use inverse-Compton constraints to show that the radiating particles
and associated magnetic field cannot provide the required pressures:
it is most likely that the `missing' pressure in these two sources is
provided by non-radiating particles such as protons.

With only two sources we cannot make a general statement, but observations of more LERGs at
relatively high redshift would be beneficial to understand this
dichotomy.

\section{Are radio galaxy clusters different from other clusters?}
\begin{figure}
\begin{center}
\includegraphics[scale=0.5,angle=0,keepaspectratio]{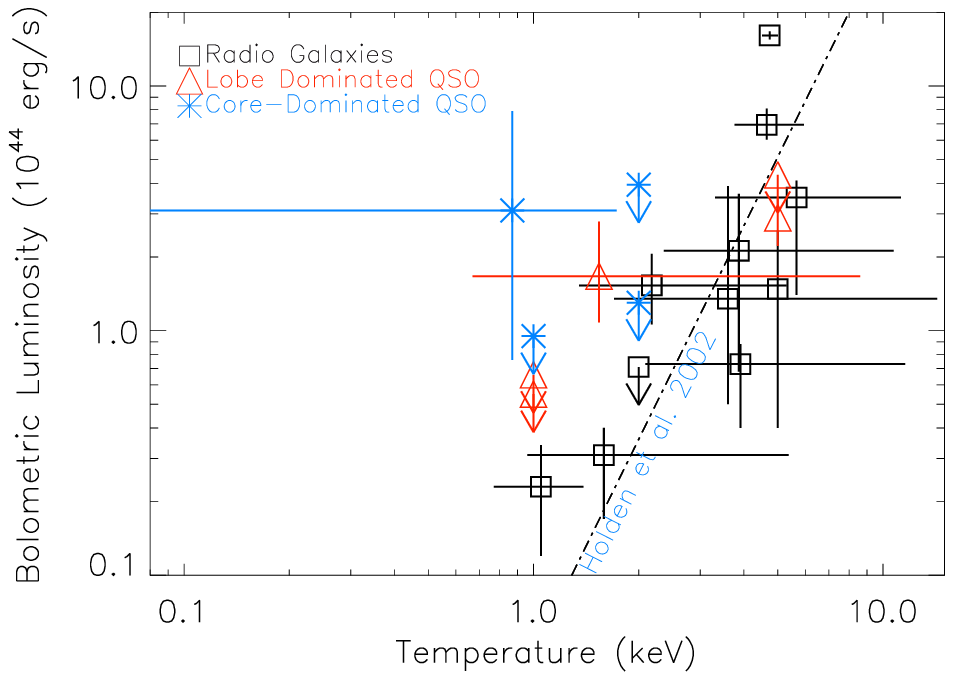}
\caption{Luminosity-Temperature relation. The line is the best fit for a sample of optically selected clusters at $z\sim0.8$; from Holden et al. (2002). Symbols are as in Fig.~\ref{fig:rel1}}
\label{fig:lxt}
\end{center}
\end{figure}

We have searched for any difference between X-ray clusters
around radio galaxies and clusters selected with a different
technique. We plot in Figure~\ref{fig:lxt} the luminosity-temperature
relation for the objects in our sample. The dashed line shows the
best-fitting correlation for an optically selected sample of clusters
from \citet{holden02}. Errors on both luminosity and temperature for
the sources in our sample are large, but we do not observe a
particular trend in the relation, and most of the sources appear in
agreement (within the uncertainties) with the $L_{\rm X}-T$ relation valid
for other clusters at the same redshift and also with the $L_{\rm
X}-T$ relation of local clusters \citep[e.g.][]{ae99} once the redshift
dependence is taken into account. We observe that most of the quasars
lie to the left of the plot. Most of the data points are upper limits,
so that it is difficult to draw strong conclusions from this
  result. However, the position of the quasars
may also indicate that the high brightness of these sources with
respect to their temperature may be due to our detection of the central
part of the cluster only. 
Since the nuclei of these
sources are brighter than those of radio galaxies, it is possible that
with the current data we are detecting only the central, denser and
cooler region of a more extended cluster, i.e. the cooling core
region.

Contrary to expectations, we do not find any source to be too hot for
its luminosity. This is in contrast with results obtained for lower
redshift radio sources, particularly in groups
\citep*[e.g.,][]{croston05a}. However the clusters we detect tend to
have relatively high temperatures and thus with the statistics
available for such high redshift sources it is not surprising that any
possible effect of heating of the ICM by the radio source is hidden or
masked by the poor statistics. On the other hand, we find that a few
sources are overluminous for their temperature. We believe that this
is more likely to be the effect of detecting only the central part of
the cluster, which may be highly luminous because of its high
density (cooler regions). Clusters at lower redshift, for which
statistics allow the separation of the cooling region, have been found
to be less luminous once the effect of the cooling flow is corrected
for \citep[e.g.][]{markevitch98,gitti07}.

\section{The environments of high-redshift sources and the
  Laing-Garrington effect}

The Laing-Garrington effect \citep{l88,g88}, in which the lobe with
the brighter or only jet appears less depolarized at GHz radio
frequencies, has been attributed since its discovery to a `hot halo'
around the powerful radio sources in which the effect was first seen.
The standard interpretation \citep{gc91} is that the counterjet side,
which in beaming models is the side further away from us, is seen
through more of the hot gas in the group or cluster environment.
As a result small-scale magnetic field variations in the hot gas give rise to
unresolved or partially resolved \citep[e.g.][]{gk05} Faraday
rotation structure and therefore `beam depolarization' in the
low-resolution images typically used to study the effect. If a $\beta$
model describes the hot gas in the system, \cite{gc91} show that the
core radius of the cluster must be comparable to the size of the radio
sources (i.e. $\sim 100$ kpc) for a significant Laing-Garrington
effect to be observed. Thus observations of the Laing-Garrington
effect essentially predict the existence of cluster-scale X-ray
emission in the source concerned. We are now in a position to compare
that prediction with the results of X-ray observations.

Laing-Garrington effect detections are only possible in sources with
detected kpc-scale jets, which represent only a fraction of the
objects in our sample. Of these, not all have depolarization
measurements in the literature. The sources with reported
Laing-Garrington effect detections in our sample are 3C\,200 (Laing
1988), 3C\,207 (Garrington, Conway \& Leahy 1991), 3C\,228 (Johnson,
Leahy \& Garrington 1995), 3C\,275.1 (Garrington \etal\ 1991) and
3C\,334 (Garrington \etal\ 1991). Of these only one (3C\,200) has an
environment with well-characterised spatial properties (Table
\ref{tab:betafit}) but these are at least qualitatively consistent
with expectations: the core radius is comparable (within the large
errors) with the $\sim 10$ arcsec lobe length. From Laing (1988),
using the results of Garrington \& Conway (1991), we can calculate the
Faraday dispersion, which we can compare to the simple model
of Garrington \& Conway (1991) using our best-fitting parameters for
the 3C\,200 environment. The ratio of the Faraday dispersions for
the measured core radius requires the source to lie at a relatively large
angle to the line of sight (see figure 7 of Garrington \& Conway
1991), which seems unlikely given its very one-sided jet (e.g. Gilbert
\etal\ 2004). However, the large errors on the core radius mean that
smaller angles can be accommodated by the data. Setting this aside,
the best-fitting parameters for the hot gas can reproduce the lobe and
counter-lobe Faraday dispersion in the Garrington \& Conway model if
the magnetic field strength in the gas is about 0.5 $\mu$G (0.05 nT).

Faraday dispersions in the other lobes with Laing-Garrington
measurements have comparable values, so we might expect roughly
similar environments for these objects. The two other sources with
cluster detections, 3C\,207 and 3C\,228, both seem to reside in
comparable environments -- 3C\,207's is rather more luminous but may
well be less centrally peaked than 3C\,200's. However, the
non-detection of environments for 3C\,334 and 3C\,275.1 is perhaps
surprising. The spectral upper limits on the luminosities of extended
emission for these sources are comparable to the detection for 3C\,200
(Table \ref{tab:spfit}), so the simplest explanation is that these
sources do have extended thermal emission that is just below our
detection threshold. There is certainly at present no gross
inconsistency between the depolarization measurements and our results.

\section{Entropy properties}
\begin{figure*}
\begin{center}
\includegraphics[scale=0.48,angle=0,keepaspectratio]{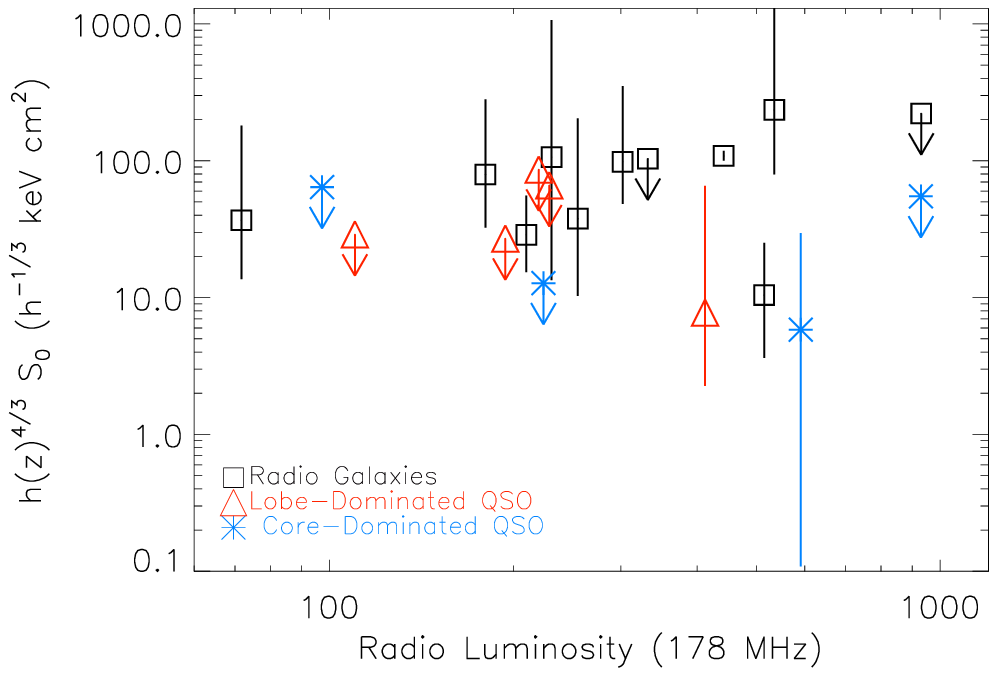}
\includegraphics[scale=0.48,angle=0,keepaspectratio]{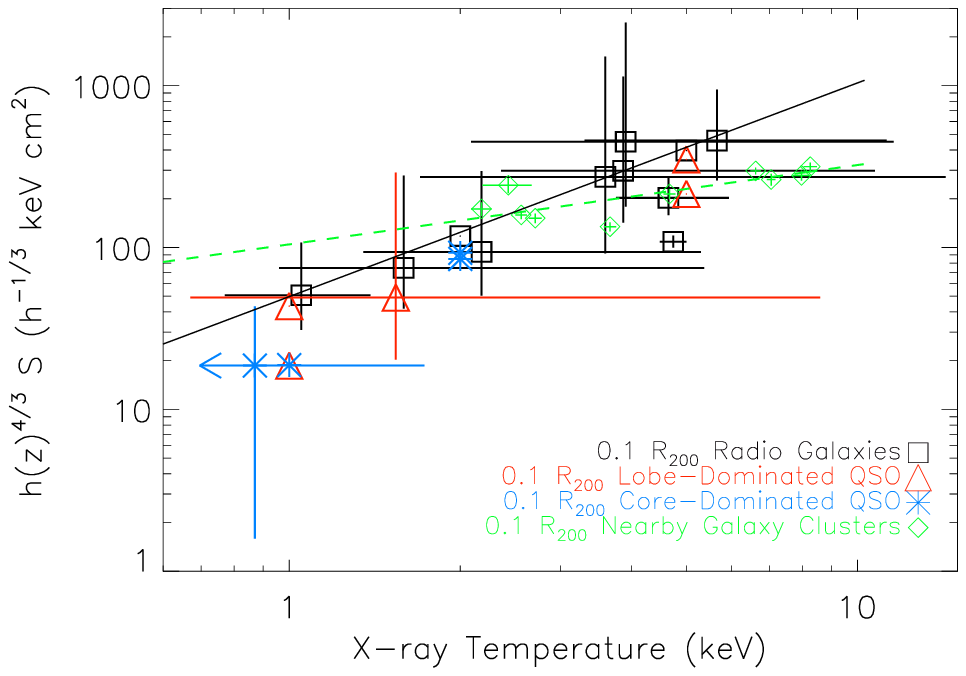}\caption{Left: Central entropy versus the radio power of the source at
178 MHz. We used the best-fitting temperature when available and
otherwise took a fixed value when the data were not of good enough quality to
perform spectral fitting. Right: Entropy calculated at 0.1 r$_{200}$
as a function of the cluster temperature. Errors are not shown when
the temperature used to fit the spectrum was a fixed parameter. (The only exception is for 3C\,309.1, where the arrow does not represent a limit but is used to show that the lower boundary of temperature error is beyond the scale of the plot.)
The red diamonds are clusters at redshift $< 0.1$ and with temperatures between 2.9 and 10 keV, from Pratt et al. (2006). The dashed line is the best fit to the cluster sample only, extrapolated to low energy. In the centre and bottom figure the entropy S has been scaled for  redshift as $S_z = h(z)^{4/3}\times S$,  where h(z) = ($\Omega_m\times(1 + z )^3 + \Omega_{\Lambda})^{0.5}$}
\label{fig:entropy}
\end{center}
\end{figure*}

Entropy is an important quantity, since it allows us to
investigate at the same time the shape of the underlying potential
well of the cluster and the properties of the ICM. Entropy is
generated in accretion shocks during the formation of the cluster
\citep[e.g.][]{tn01,voitetal03,voit05,borganietal05}, but it can be
modified by non-gravitational internal processes after the formation
of the cluster itself. Much emphasis has been placed on entropy in
studies of galaxy clusters, from both a theoretical \citep[e.g.][ and
references therein]{voitetal02,voitetal05,muanwong06} and
observational \citep[e.g.][]{tjpetal03,gwpma05,piffaretti05,gwp06}
perspective. Recent results show that entropy is higher than expected
from pure gravitational models on cluster spatial scales out to
at least half the virial radius \citep*[e.g.][ and references
therein]{gwp06} and not only in the central regions. Although the
pre-heating scenario \citep[e.g.,][]{kaiser91,
evrhen91,valageassilk99} is now considered unlikely
\citep*[e.g.][]{tjpetal03,gwpma03,gwpma05}, mechanisms such as heating
and cooling due to more recent supernovae or AGN activity are still a viable
explanation for the general excess of entropy. There is a
consensus about the significant effect of feedback from a central AGN
in the evolution of the ICM. An important result is that central AGNs
can act not only in the central region of what were historically
called ``cooling flow'' clusters, but also on large scales, e.g.
through sonic shocks, observed as ripples in the ICM
\citep{acfperseus} or strong shock waves \citep{formanm87}. This is
particularly relevant as the AGN hypothesis to explain the excess
entropy in galaxy clusters can also apply to the large scale excess
and it is not limited to the central entropy excess initially detected
in cool systems \citep[e.g.][]{tjpetal99,Lldaviesetal00,tjpetal03}.

These recent results on entropy motivated us to investigate the
entropy properties of the clusters in our sample. Although the errors are
large, this is the first time that entropy has been investigated
for the ICM around radio sources at this redshift. Here we adopt the
accepted definition of entropy in the X-ray:
\begin{equation}
S = kT/n_e^{2/3}
\end{equation}

\noindent In order to compare with low-redshift clusters, we scale this with
redshift using the relation $S_z = h^{4/3}(z) \times S$. For the
sources in the sample we are unable to produce entropy profiles as we
only have a global temperature and therefore the shape of the profile would
depend exclusively on the shape of the density profile. We thus only
obtained global entropy properties at specific distances from the
centre of the cluster.

Figure~\ref{fig:entropy}, left, shows the central entropy $S_{\rm 0}$
compared to the radio power at 178 MHz. The isotropic radio power is
often taken to be a proxy of the jet power \citep{willott99}, and thus
should be correlated with the power being injected into the ICM. The entropy reflects the accretion history of the cluster but also the
influence of non-gravitational processes. We find that the central
entropies of all the objects in the sample are similar and there is no
correlation between $S_0$ and $L_{\rm 178 MHz}$. We also see that RGs
and quasars are scattered equally in the space defined by these
two parameters, although caution should be used as regards the CDQs since their low-frequency radio emission may be
contaminated by emission from the jet. The lack of correlation between the two quantities
may be the result of the short life time of the radio source with
respect to the cluster age; the two quantities may be unrelated
because they arise from processes that take place at different
epochs.

In the right-hand panel of Figure~\ref{fig:entropy} we plot the
entropy at 0.1 $r_{\rm 200}$ versus the temperature of each cluster
(upper limits are not shown for clarity, but symbols without error
bars should be interpreted as limits). This is similar to what
\citet*{gwp06} and others have done for lower redshift clusters. In the Figure the
objects in our sample are compared to a sample of 10 clusters at
redshift less than 0.2 \citep*{gwp06}.  We also plot the 
  best-fitting temperature-entropy relation
for this low-$z$ cluster sample, extrapolated to 
low temperatures. We observe that, at 0.1~$r_{\rm 200}$,  clusters around 
RGs with temperature above 2.5 keV appear similar in their entropy 
properties, within the uncertainties, to clusters at
lower redshift. However they are preferentially found above the 
best-fitting temperature-entropy relation for local clusters, suggesting that they have a larger amount of entropy than their local counterpart.
The cooler radio-galaxy clusters all lie below the 
local $S-T$ relation of Pratt \etal, and although we do not show it 
here, our data cannot be fitted by the best-fit $S-T$ relation of 
slope 0.65 found by  \citet*{tjpetal03} using a sample of local 
groups, galaxies and clusters, which may be a more appropriate comparison for the low-temperature systems in our sample.
We have calculated
the entropy at 0.5~$r_{\rm 200}$ and the trend is similar. However, in
many cases calculation of the density at this radius requires a
  large extrapolation from the data so that the results may be biased.

Using the {\sc asurv} package \citep{isobe90asurv} to account for
censored data, we find that  the relationship between $S$ (0.1 $r_{\rm 200}$) and $T$  is fitted by a
relation of slope 1.32$\pm0.22$, in reasonable agreement with what is expected from self-similar models 
(slope of 1; \citealt*{tjpetal99}) -- note that errors here are 1$\sigma$.  
This seems to suggest that there is 
less entropy in high-redshift cool systems than their local counterparts, which is the opposite of what is
expected if, as observed for larger sample of galaxy groups and
clusters, the cool systems are more strongly affected by feedback processes. It
also appears counter-intuitive given the presence of a powerful radio
source at the centre of all our clusters. However, the presence of the
radio source may represent a concrete explanation for this behaviour.
The AGN (at least for the conventional AGN that dominates our
sample) needs to be fuelled to be activated and for the accretion to be radiatively efficient, as is the case for most of the sources in the sample \citep*[see][]{bel06}, it needs the accreted gas to be relatively cold. The required cool
gas becomes available when low-entropy gas is present close enough to
the accretion region. For this to happen, it is commonly accepted that
the cluster must be in a virialized state. Since the cluster requires a
time which is much longer than the lifetime of synchrotron emitting
electrons at 1.4 GHz ($\sim10^8$ yrs) to relax, the radio sources in
this sample may be in a state in which the energy of the outburst due
to the AGN activation has not yet been transferred to the ICM, even in
the central region. Clusters at lower redshift, such as those in
\citet*{gwp06} or groups \citep*{tjpetal03} have had more time for the
distribution of AGN energy output throughout the cluster, and their
high entropy level, especially within 0.1 $r_{\rm 200}$, may be the effect of repeated outburst. 
If we accept the interpretation of cyclic activation of the central supermassive black hole, 
the duty cycle must be higher in low-redshift clusters. 
It is possible in our present sample that we are witnessing the first
activation of a radio galaxy at the centre of a dense cluster-like
environment, particularly bearing in mind that the 3CRR sources
on which our sample is based are at the top end of the luminosity
function and are rare at all epochs. This may also be interpreted 
as suggesting that the duty cycle of more luminous radio sources is much lower than that of less active AGN.
In other words, the sources in the sample with low central entropy may
be selected to be those in which the effect of the radio source 
 is not yet observable in entropy, but will be in the future.

If this result can be confirmed with better
statistics, and/or the larger samples that will be available with
X-ray instruments such as {\it XEUS}, it may give important constraints
on the overall mechanism of AGN heating of the ICM, entropy distribution
in large scale structure and, last but not least, the formation,
evolution and ageing of radio sources in dense environments.

\section{Conclusions}
We have analysed X-ray data from \cha\ and \xmm\ of 20 powerful RGs and quasars at redshift $0.45<z<1$ with the aim of investigating their external environment and its interaction with the bright radio source. We find that: 
\begin{enumerate}
\item RGs and quasars inhabit a range of environments without any obvious
  relationship to radio properties such as radio power and size.

\item more than 60 per cent of the sources lie in X-ray emitting environments of luminosity greater than $10^{44}$ \es. The poorest environments in this redshift range are consistent with
  moderate-luminosity groups ($\Lx \sim 10^{42}$ \es), the richest with Abell class 2 ($\Lx>5 \times10^{44}$ \es)) or above clusters.

\item Our results are in agreement with unification schemes that predict that the environment of sources oriented with various angles to the line of site should be isotropic. Most of the objects whose jet is oriented close to the plane of the sky  (RGs) are detected to lie in rich environments while for 6 out of 9 quasars we are able to estimate upper limits for the existence of a cluster-like component around them. These upper limits occupy the same parameter space of luminosities as the detected sources. Seven of the sources are found to be point-like. However three of them are core-dominated quasars for which the detection of a low-surface brightness X-ray component is hampered by the the bright X-ray core emission.

\item  Within the uncertainties, these are normal environments, with no evidence that they deviate from the temperature/luminosity relationship observed in low-$z$ normal clusters.
  Therefore there is no strong evidence that the presence of a radio source
  requires a peculiar environment or strongly affects the cluster. This may also suggest that deviation from self similarity and the scatter around scaling laws is likely to be affected by other processes such as mergers \citep{maughan07}, rather than the interaction between the radio source and the central cluster.

\item For the first time for radio active sources at this redshift we investigate the entropy of the ICM. To our surprise we find that the entropy/temperature relation is steeper than that observed for low-redshift normal galaxy clusters, and closer to the relation predicted by self-similar models. We argue that this is the result of processes occurring at different epochs, the intensity of the AGN duty cycle at low and high redshift, and that the heating from the radio source outburst may have not have had the time yet  to be observed in the ICM entropy for sources at redshift above 0.5.

\item Most of the sources appear close to pressure balance with the cluster, with the exception of the two LERGs in the sample which appear under-pressured. Although measurements of the internal pressure are still uncertain, this comparison is more precise than using simple minimum pressures from radio data.  The result on LERGs may be the first evidence of the presence of non-radiating particles that contribute to the radio source internal pressure in powerful radio sources.

\end{enumerate}

\appendix

\section{Notes on individual sources}\label{app:A1}
In this Appendix we describe details of the source-specific analysis
for each of the sources in the sample. For each source we show
wavelet-decomposed images in the soft (0.5-2.0 keV) and hard (2.5-7.0
keV) energy bands, together with the radial profile fitting described
in the text, with both a PSF alone and a PSF plus $\beta$-model when
required. The fitting results for radial profile and spectral analysis
are in Tables~ \ref{tab:betafit} and ~\ref{tab:spfit}. The images and profile can be inspected in Fig. A1-A18.
The contours in the images are from VLA data at 1.4 GHz,  except where
otherwise stated. No radio overlays are shown for the 4 CDQs due
to the small size of the source  (see also Sec. 3.2).

\begin{itemize}
\item {\bf 3C\,6.1:} The radial profile is well fitted with a PSF only and there is no significant improvement in the fit if an extra
component is added. We fix the $\beta$ parameter to 0.5 and the core
radius $r_{\rm c}$ to 15 arcsec = 115 kpc, for a 2 keV cluster, which
is a typical \rc\ value for a cluster in the local universe ($\sim
125$ kpc). We calculate a 3$\sigma$ upper limit for any cluster-like
extended emission with these parameters.

\begin{figure*}
\begin{centering}
\hfill
\hfill
\caption{From left to right: Soft (0.5-2.0 keV) and hard (2.5-7.0 keV) wavelet-decomposed image for 3C\,6.1. Contours are from a 1.4-GHz radio map (A configuration, L band) and are logarithmically spaced. Radial profile fitted with a simple PSF. The $\beta$-model fitting of the radial profile is not shown as only an upper limit was calculated.}
\label{fig:3c6.1}
\end{centering}
\end{figure*}

We extracted the spectrum in the same source and background area as
used for the radial profile. We fitted the spectrum with 1) a power law
absorbed by Galactic absorption and 2) a thermal model. We find a similar goodness of fit.
The power-law slope is found to be $1.7\pm0.5$ when the spectrum is fitted between 0.5-2.5 keV and
$1.4\pm0.23$ if the 0.5-6.0 keV band is used. Both are consistent
with the spectrum being due to the wings of the PSF. However, the
number of counts does not correspond to what is expected from the
wings of the PSF at the distance from the centre we use (a circle of radius 5-arcsec
 was used to mask the central unresolved point source). The
source is only slightly piled-up. The other option for the nature of
this extended emission may be IC from the
region coincident with the radio lobes or the hotspots. However those
areas were masked prior to spectrum extraction, and \citet{croston05b}
find only upper limits for the IC emission from the lobes. For this
reason, we evaluated the luminosity of the emission detected in the circle of radius 
49.2 arcsec used for the spectral analysis by fixing a temperature of
2 keV and assuming it to be thermal. Images of 3C\,6.1 are in
Figure~\ref{fig:3c6.1}.

\noindent \item {\bf 3C\,184}
This source was analysed in detail by \citet{bel04}. The point source is not spatially separated from the extended emission with the \xmm\ data, and 20 ks are not sufficient to detect cluster emission with \cha. However, the \xmm\ spectrum is best fitted with 3 components, one of which is a \mekal\ with at best-fit temperature of $\sim$3.6 keV. Images and profiles can be inspected in \citet{bel04}.

\noindent \item {\bf 3C\,200:}
This is a low-excitation radio galaxy, so the core is faint with respect to the lobe emission, making the investigation of diffuse emission easier. The statistics are good enough to constrain the radial profile and spectral parameters. Figure~\ref{fig:3c200} shows images for this source.

\begin{figure*}
\begin{centering}
\hfill
\hfill
\caption{From left to right: Soft (0.5-2.0 keV) and hard (2.5-7.0 keV) wavelet-decomposed image for 3C\,200. Contours are from a 1.4-GHz radio map (B configuration, L band) and are logarithmically spaced. Radial profile fitted with simple PSF and a PSF plus a $\beta$-model.}
\label{fig:3c200}
\end{centering}
\end{figure*}

\noindent \item {\bf 3C\,207:} There are a total of 190 net counts in
the region used for the spectral analysis: a circle of radius 15
arcsec with the exclusion of a circle of radius 5 arcsec to mask the
core, and areas to mask the western lobe and jet. The spectrum can be
fitted with a simple power-law model of slope 1.15, giving a
$\chi^2$/d.o.f = 3.06/4.  There is no statistically significant need
for a thermal component in agreement with \citet{brunetti02} and \citet{gambill03}.
 However, the radial profile analysis strongly
suggests the presence of an extended component with a King profile
distribution and standard parameters. We thus calculated the fraction
of counts from the wings of the PSF expected in the area used for the
spectral analysis, and we derived the normalisation of the power law
model in this area. We then fixed the value of $\Gamma$  to
be equal to the lower limit of $\Gamma$ found while fitting the
core spectrum. This is justified by the flattening of a point source
spectrum in the PSF wings. Finally we assumed that the rest of the
counts were from extended, cluster-like emission that was fitted with
a \mekal\ model of fixed temperature 5 keV. Given the uncertainties of
the spectral fitting, we have assigned the source a quality flag
of D (detection only) despite the fact that $\beta$ model parameter
are well constrained. These results should thus only be regarded as
giving one plausible scenario. Using the same data \citet{brunetti02} interpret most of
the X-ray emission surrounding one of the lobes as non-thermal;
however, they do not exclude the possibility that a low-surface
brightness cluster may be present.

\begin{figure*}
\begin{centering}
\hfill
\hfill
\caption{From left to right: Soft (0.5-2.0 keV) and hard (2.5-7.0 keV) wavelet-decomposed image for 3C\,207. Contours are from a 1.4-GHz radio map (A configuration, L band) and are logarithmically spaced. Radial profile fitted with simple PSF and a PSF plus $\beta$-model.}
\label{fig:3c207}
\end{centering}
\end{figure*}

\noindent \item {\bf 3C\,220.1:} 
This source was detected previously with {\em ROSAT} \citep{mjh98a} and using the same \cha\ data by \citet{dmw01}. It is one of the best examples of a rich cluster around a radio galaxy in this redshift range, and the precision of physical parameters obtained from spatial and spectral analysis are comparable to X-ray clusters in the local universe. Our results are in perfect agreement with previous studies of this cluster.

\begin{figure*}
\begin{centering}
\hfill
\hfill
\caption{From left to right: Soft (0.5-2.0 keV) and hard (2.5-7.0 keV) wavelet-decomposed image for 3C\,220.1. Contours are from a 1.4-GHz radio map (B configuration, L band) and are logarithmically spaced. Radial profile fitted withe simple PSF and a PSF plus $\beta$-model.}
\label{fig:3c220.1}
\end{centering}
\end{figure*}

\noindent \item {\bf 3C\,228:}
Two exposures for this source give a total of $\sim24$ ks of flare-clean \cha\ observation. This allows us to explore the $\beta$-$r_{\rm c}$ parameter space but not to constrain the two parameters. The point source dominates the X-ray emission out to 6 arcsec and this was excluded prior to spectrum extraction. We thus fixed the $\beta$ and $r_{\rm c}$ to 0.5 and 15 arcsec respectively and only calculated errors on the normalisation. 
\begin{figure*}
\begin{centering}
\hfill
\hfill
\caption{From left to right: Soft (0.5-2.0 keV) and hard (2.5-7.0 keV) wavelet-decomposed image for 3C\,228. Contours are from a 1.4-GHz radio map (B configuration, L band), and are logarithmically spaced. Radial profile fitted withe simple PSF and a PSF plus $\beta$-model.}
\label{fig:3c228}
\end{centering}
\end{figure*}

The spectrum is well fitted with a \mekal\ model of $kT= 3.9$ keV and
metallicity 0.3 solar. However it can also be fitted with a power law
of $\Gamma= 1.8$. The two fitted models are statistically
equivalent; however, the nucleus of the source has a flatter
($\Gamma=1.6\pm0.1$) spectrum, so that this emission is unlikely to
come from the wings of the PSF. IC emission from the radio lobes may
be the other source of uncertainty but \citet{croston05b} only find
upper limits on IC flux from the radio lobes. We thus assume that a
thermal model describing the external environment is the most likely
interpretation for the X-ray emission detected by our spectral
analysis.

\noindent \item {\bf 3C\,254:} The radial profile of 3C\,254 is
dominated by the PSF out to 40 arcsec. This is one of the
significantly piled up sources, with a pileup fraction of 20 per cent.
However, for both the radial profile and the spectral analysis, the PSF
alone is not sufficient to account for all the X-ray emission and we
constrain the shape ($\beta$ model) and spectral parameters. From the
best fit \rc\ value the object appears rather compact and the low
temperature we find may suggest that we have possibly detected the
galaxy atmosphere rather than a cluster.
\begin{figure*}
\begin{centering}
\hfill
\hfill
\caption{From left to right: Soft (0.5-2.0 keV) and hard (2.5-7.0 keV) wavelet-decomposed image for 3C\,254. Contours are from a 1.4-GHz radio map (configuration A, band L) and are logarithmically spaced. Radial profile fitted withe simple PSF and a PSF plus $\beta$-model.}
\label{fig:3c254}
\end{centering}
\end{figure*}
\citet{cf03} detect extended X-ray emission around 3C\,254 with a luminosity in the 0.5-7 keV range of $(3.1\pm1.2)\times10^{43}$ \es, which is in agreement with what we find within the errors.

\noindent \item {\bf 3C\,263:} 
The source is heavily piled up. We used the piled up PSF model to characterise the point source distribution. However, given the uncertainties of this model for heavily piled up sources, we cannot account for all the residuals and the fit is not good. The addition of a $\beta$-model improves the fit but the parameters we find are not physical, and suggest a best-fit value for the core radius close to zero. We interpret this result as the effect of our poor modelling of the wings of the PSF for a heavily piled up source. As a result, we fix $\beta$ and \rc\ to canonical values and obtained upper limit only from both the radial profile and spectrum. 
\begin{figure*}
\begin{centering}
\hfill
\hfill
\caption{From left to right: Soft (0.5-2.0 keV) and hard (2.5-7.0 keV) wavelet-decomposed image for 3C\,263. Contours are from a 1.4-GHz radio map (configuration A, L band) and are logarithmically spaced. Radial profile fitted with a simple PSF.}
\label{fig:3c263}
\end{centering}
\end{figure*}

\citet{cf03} and \citet{mjh02} previously studied this source and they
both claimed detection of cluster emission, albeit of low surface
brightness. However their modelling of the PSF did not take pileup
into account and their result may be due to pollution from the
point source, although 3C\,263 seems to lie in a spectroscopically
  confirmed optical cluster \citep{hall95}. To summarize, it may be possible that a cluster exists around 3C\,263, but given the poor fitting results we obtain we adopt the safe solution and use upper limits only for this source.

\noindent \item {\bf 3C\,265:} The upper value of the $\beta$
parameter was not constrained, which can be explained as the result of
the poor statistics for this object, mostly due to the large mask used
to exclude the radio lobes. The relatively low temperature we measure
suggests that the source lies in a group or that we are detecting only
the very centre of a larger cluster which may have a cool core.

\begin{figure*}
\begin{centering}
\hfill
\hfill
\caption{From left to right: Soft (0.5-2.0 keV) and hard (2.5-7.0 keV) wavelet-decomposed image for 3C\,265. Contours are from a 1.4-GHz radio map (configuration B, L band), and are logarithmically spaced. Radial profile fitted with a simple PSF and a PSF plus $\beta$-model.}
\label{fig:3c265}
\end{centering}
\end{figure*}

\item {\bf 3C\,275.1} The radial profile and the spectral analysis
both indicate that the X-ray emission is from a point source. We thus
carried out upper limit calculation for the existence of a cluster
component. This is not in agreement with what was found by \citet{cf03}
who detected a cluster of bolometric X-ray luminosity
$7.6\pm0.9\times10^{43}$ \es.

\begin{figure*}
\begin{centering}
\hfill
\hfill
\caption{From left to right: Soft (0.5-2.0 keV) and hard (2.5-7.0 keV) wavelet-decomposed image for 3C\,275.1. Contours are from a 1.4-GHz radio map (A configuration), and are logarithmically spaced. Radial profile fitted with simple PSF.}
\label{fig:3c275.1}
\end{centering}
\end{figure*}

\noindent \item {\bf 3C\,280:} The $\beta$-model parameters are not
constrained. However, there is a significant improvement over a PSF
model alone if an extended model is added. We use here the
best-fitting parameters of the $\beta$ model to characterise the
extended emission: the derived values should therefore be used with
caution. Similarly, the count statistics of the current \cha\
observation do not allow us to constrain the temperature, although the
spectrum is better fitted with a thermal model than a power law. The luminosity we measure is within the upper limit of \citet{ddh03}, although
we claim a detection. 

\begin{figure*}
\begin{centering}
\hfill
\hfill
\caption{From left to right: Soft (0.5-2.0 keV) and hard (2.5-7.0 keV)
  wavelet-decomposed image for 3C\,280. Contours are from a 1.4-GHz
  radio map (A configuration), and are logarithmically spaced. The radial profile is fitted with a simple PSF and a PSF plus $\beta$-model.}
\label{fig:3c280}
\end{centering}
\end{figure*}

 \item {\bf 3C\,292:} \citet{bel04} give a detailed analysis of this source.

 \noindent \item {\bf 3C\,295:} This rich cluster was previously
  studied by \citet{allen01-3c295} using a \cha\ ACIS-S observation.
  We have used a more recent 90-ks observation made with the ACIS-I
  detector. These data have extremely good statistics and would allow
  a much more detailed analysis of the source than we have
  presented here. For consistency, however, we have analysed 3C\,295 in the same way as all the
  other sources. The host cluster of 3C\,295 is the most luminous
  cluster around a 3CRR source in the redshift range of the paper. Our
  morphological analysis agrees with what was found by Allen \etal
  Since we calculate a global temperature for the cluster, we find a
  slightly lower temperature than Allen \etal, who exclude the cooling
  flow region. For the same reason, 3C\,295 appears too luminous for
  its temperature (see Fig.~\ref{fig:lxt}). However, if we accept the
  measurements of Allen \etal beyond 50 kpc, the source is found more
  in line with the $L_{\rm X}-T$ relation shown in Fig.~\ref{fig:lxt}.
  Despite being an extreme case in our sample (because of the number
  of X-ray counts detected and high luminosity), 3C\,295 appears
  similar to the other 3CRR sources in the 0.45--1 redshift range when
  its global properties are considered.

\begin{figure*}
\begin{centering}
\hfill
\hfill
\caption{From left to right: Soft (0.5-2.0 keV) and hard (2.5-7.0 keV)
  wavelet-decomposed image for 3C\,295. Contours are from an 8.6-GHz
  radio map (A configuration), and are logarithmically spaced. The radial profile is fitted with a simple PSF and a PSF plus $\beta$-model.}
\label{fig:3c295}
\end{centering}
\end{figure*}

\noindent \item {\bf 3C\,309.1:} This is the only core-dominated
quasar for which we have been able to estimate the extent of the
thermal emission. The source was observed with a small window and thus
only the area covered by the detector was used. A King profile in
addition to the PSF is required to fit the radial profile of 3C\,309.1,
and the shape of the profile is constrained. In order to increase the
S/N ratio we reduced the area for extraction of the spectrum to a
circle of radius 32 arcsec (instead of r = $\sim40$ arcsec used for
the radial profile). The bright point source was masked by a circle of
radius 9 arcsec, thus reducing considerably the contamination of the
PSF, even for a piled up source. The readout streak was also masked.
The low number of counts in this region (60 net counts) requires the
use of Cash statistics. Although a power law gives an acceptable fit
for the spectrum, the slope is too steep ($\Gamma=2$), and it
is not in agreement with the spectrum of the point source. The
spectrum of 3C\,309.1 is better fit with a \mekal\ model of low
temperature $\sim0.9$ keV, even though the statistical errors are
large. Therefore, spatial and spectral data both seem to indicate that this
object lies in a X-ray emitting external environment. With the current
data we cannot say whether our detection corresponds to a galaxy
group, a bright elliptical galaxy or the centre of a cooling core
cluster atmosphere and additional observations of this source would
help to shed light on the nature of its environment.

\begin{figure*}
\begin{centering}
\hfill
\hfill
\caption{From left to right: Soft (0.5-2.0 keV) and hard (2.5-7.0 keV) wavelet-decomposed image for 3C\,309.1.  Radial profile fitted with a simple PSF and a PSF plus $\beta$-model.}
\label{fig:3c309.1}
\end{centering}
\end{figure*}

\noindent \item {\bf 3C\,330:} The \cha\ data allow us to fully
determine the spatial and spectral characteristics of this source. Our
results are in agreement with those of \citet{mjh02}.
\begin{figure*}
\begin{centering}
\hfill
\hfill
\caption{From left to right: Soft (0.5-2.0 keV) and hard (2.5-7.0 keV) wavelet-decomposed image for 3C\,330. Contours are from a VLA 1.4-GHz radio map taken in configuration A,  and are logarithmically spaced. Radial profile fitted with a simple PSF and a PSF plus $\beta$-model.}
\label{fig:3c330}
\end{centering}
\end{figure*}

\noindent \item {\bf 3C\,334:} 
The radial profile is well modelled with a PSF accounting for pileup. The spectral distribution suggests that if a thermal component is present, it is relatively cold as the 20 net counts left after excluding the core emission are below 1 keV. We thus fixed the temperature to be 1 keV and calculated a 3$\sigma$ limit for the presence of an extended environment.

\begin{figure*}
\begin{centering}
\hfill
\hfill

\caption{From left to right: Soft (0.5-2.0 keV) and hard (2.5-7.0 keV) wavelet-decomposed image for 3C\,334. Contours are from a 1.4-GHz radio map taken in configuration B, and are logarithmically spaced. Radial profile fitted with a piled up model of the PSF.}
\label{fig:3c334}
\end{centering}
\end{figure*}

\noindent \item {\bf 3C\,345:} 
The radial profile is modelled with a PSF, accounting for pileup. We fix the temperature of any thermal component to 2 keV to estimate the upper limit of thermal emission from the spectrum.

\begin{figure*}
\begin{centering}
\hfill
\hfill
\caption{From left to right: Soft (0.5-2.0 keV) and hard (2.5-7.0 keV) wavelet-decomposed image for 3C\,345. Radial profile fitted with a piled up model of the PSF.}
\label{fig:3c345}
\end{centering}
\end{figure*}

\noindent \item {\bf 3C\,380:} Like 3C\,345, 3C\,380 is a
core-dominated quasar for which the modelling of the PSF is
complicated by pileup. The radial profile is well fitted with a
PSF model and upper limits are evaluated for the existence of an
extended thermal component using our standard values for $\beta$ and \rc\
and assuming a temperature of 2 keV.
\begin{figure*}
\begin{centering}
\hfill
\hfill
\caption{From left to right: Soft (0.5-2.0 keV) and hard (2.5-7.0 keV) wavelet-decomposed image for 3C\,380.  Radial profile fitted with a piled up model of the PSF.}
\label{fig:3c380}
\end{centering}
\end{figure*}

\noindent \item {\bf 3C\,427.1:} This source is a LERG with a very
weak nucleus. The emission extends out to a radius of 200 kpc and is
thermal in nature, with a best-fit temperature of 5.7 keV.
Using the radial profile we can only map the extended emission out to
25 per cent of the virial radius and there may be some indication that a secondary $\beta$ model is needed to fit the profile, but its
parameters are not constrained. \citet{croston05b} did not detect
significant counts from the radio lobes. Nevertheless, we applied a conservative approach and excluded the spatial
areas coincident with the radio lobes when extracting both spectrum
and radial profile. The thermal model fit gives an interestingly high
temperature for a source at this redshift. The Fe K$_{\alpha}$ line
appear very strong. If the chemical abundance is left as a free
parameter, the best fit give $Z/Z_{\odot}$ = 1.68, and 1$\sigma$
errors suggest that it is  a greater than 0.76 (although the upper
limit is found to be 3.8. We thus decided to fix $Z/Z_{\odot}$ to 1
and fit with this parameter frozen. A power-law model gives a worse
fit but not significantly so ($\Delta\chi^2 = 2.5$) and a best fit of
$\Gamma = 1.6\pm0.2$.

\begin{figure*}
\begin{centering}
\hfill
\hfill
\caption{From left to right: Soft (0.5-2.0 keV) and hard (2.5-7.0 keV) wavelet-decomposed image for 3C\,427.1. Contours are from a 1.4-GHz radio map (Configuration A), and are logarithmically spaced. Radial profile fitted with a simple PSF and a PSF plus $\beta$-model.}
\label{fig:3c427.1}
\end{centering}
\end{figure*}

\noindent \item {\bf 3C\,454.3:} Even though the radial-profile fit
using a piled-up PSF shows significant residuals, we interpret this as
the effect of a poor modelling of the pileup rather than the detection
of a secondary component. As a result, for this core-dominated
quasar we calculate 3$\sigma$ upper limits for the presence of an
extended and thermal (1 keV) environment.
\begin{figure*}
\begin{centering}
\hfill
\hfill
\caption{From left to right: Soft (0.5-2.0 keV) and hard (2.5-7.0 keV) wavelet-decomposed image for 3C\,454.3. Radial profile fitted with a piled up model of the PSF.}
\label{fig:3c454.3}
\end{centering}
\end{figure*}

\end{itemize}
\section*{Acknowledgements}
We are grateful to Gabriel W. Pratt for enlightening discussion about entropy and to Alexey Vikhlinin for providing his wavelet algorithm. We thank the anonymous referee for useful and detailed comments that improved the manuscript. E.B. thanks PPARC for support and MJH thanks the Royal Society for a
research fellowship. This work is
  partly based on observations obtained with {\it XMM-Newton}, an
ESA science mission with instruments and contributions directly funded
by ESA Member States and NASA. This research has made use of the SIMBAD
database, operated at CDS, Strasbourg, France, and the NASA/IPAC
Extragalactic Database (NED) which is operated by the Jet Propulsion
Laboratory, California Institute of Technology, under contract with
the National Aeronautics and Space Administration.

\end{document}